\documentstyle[12pt,epsf,geompsfi]{article}
\newcommand{\nc}{\newcommand}
\nc{\renc}{\renewcommand}

\renc{\baselinestretch}{1}
\nc{\com}[1]{\ \\ \ {\bf \# {#1}}\\ \ }
\nc{\bort}[1]{}
\newlength{\overeqskip}
\newlength{\undereqskip}
\nc{\be}[1]{\begin{equation} \mbox{$\label{#1}$}}
\nc{\bea}[1]{\begin{eqnarray} \mbox{$\label{#1}$}}
\nc{\Section}[2]{\section{\sc #2}\label{#1}}
\nc{\Subsection}[2]{\subsection{\sc #2}\label{#1}}
\nc{\Bibitem}[1]{\bibitem{#1}}
\nc{\Label}[1]{\label{#1}}
\nc{\eea}{\vspace{\undereqskip}\end{eqnarray}}
\nc{\ee}{\vspace{\undereqskip}\end{equation}}
\nc{\bdm}{\begin{displaymath}}
\nc{\edm}{\end{displaymath}}
\nc{\dpsty}{\displaystyle}
\nc{\bc}{\begin{center}}
\nc{\ec}{\end{center}}
\nc{\ba}{\begin{array}}
\nc{\ea}{\end{array}}
\nc{\bab}{\begin{abstract}}
\nc{\eab}{\end{abstract}}
\nc{\btab}{\begin{tabular}}
\nc{\etab}{\end{tabular}}
\nc{\bit}{\begin{itemize}}
\nc{\eit}{\end{itemize}}
\nc{\ben}{\begin{enumerate}}
\nc{\een}{\end{enumerate}}
\nc{\bfig}{\begin{figure}}
\nc{\efig}{\end{figure}}
\nc{\seqnoll}{\setcounter{equation}{0}}
\renc{\theequation}{\thesection.\arabic{equation}}
\renc{\Section}[2]{\bc\section{\sc #2}\Label{#1}\seqnoll\ec}

\nc{\refc}[1]{\mbox{Ref.~\cite{#1}}}
\nc{\refs}[1]{\mbox{Refs.~\cite{#1}}}

\nc{\eqs}[2]{\mbox{Eqs.~(\ref{#1},\,\ref{#2})}}
\nc{\eq}[1]{\mbox{Eq.~(\ref{#1})}}

\nc{\figs}[2]{\mbox{Figs.~\ref{#1} and \ref{#2}}}
\nc{\fig}[1]{\mbox{Fig.~\ref{#1}}}

\nc{\figcap}[1]{\refstepcounter{figure}
        {\bf Figure \thefigure}: {\small\sl #1}}
\nc{\tabcap}[1]{\refstepcounter{table}
        {\bf Table \thetable}: {\small\sl #1}}

\def\jump{\vskip 1truecm\noindent}

\nc{\tag}[1]{\label{#1} \marginpar{{\footnotesize #1}}}
\nc{\mtag}[1]{\label{#1} \mbox{\marginpar{{\footnotesize #1}}}}

\nc{\etal}{\mbox{\it et al. }}
\nc{\ie}{{\it i.e.}}
\nc{\eg}{{\it e.g.}}

\nc{\arreq}{&\!\!\!=\!\!\!&}
\nc{\arrmi}{&\!\!\!!-\!\!\!&}
\nc{\arrpl}{&\!\!\!+\!\!\!&}
\nc{\arrap}{&\!\!\!\approx\!\!\!&}
\nc{\non}{\nonumber}
\nc{\align}{\!\!\!\!\!\!\!\!&&}

\nc{\mat}[4]{{\left(\ba{cc} #1 & #2 \\ #3 & #4 \ea\right)}}

\def\simleq{\; \raise0.3ex\hbox{$<$\kern-0.75em
      \raise-1.1ex\hbox{$\sim$}}\; }
\def\simgeq{\; \raise0.3ex\hbox{$>$\kern-0.75em
      \raise-1.1ex\hbox{$\sim$}}\; }
\nc{\DOT}{\hspace{-0.08in}{\bf .}\hspace{0.1in}}
\nc{\Laada}{\hbox {$\sqcap$ \kern -1em $\sqcup$}}
\nc\loota{{\scriptstyle\sqcap\kern-0.55em\hbox{$\scriptstyle\sqcup$}}}
\nc\Loota{{\sqcap\kern-0.65em\hbox{$\sqcup$}}}
\nc\laada{\Loota}
\nc{\qed}{\hskip 3em \hbox{\BOX} \vskip 2ex}

\def\Im{{\rm Im}\hskip2pt}
\nc{\real}{{\rm I \! R}}
\nc{\Z}{{\sf Z \!\!\! Z}}
\nc{\complex}{{\rm C\!\!\! {\sf I}\,\,}}
\def\bigid{\leavevmode\hbox{\small1\kern-3.8pt\normalsize1}}
\def\id{\leavevmode\hbox{\small1\kern-3.3pt\normalsize1}}
\nc{\slask}{\!\!\!\!/}
\nc{\sslask}{\!\!\!/}
\nc{\bis}{{\prime\prime}}
\nc{\pa}{\partial}
\nc{\na}{\nabla}
\nc{\ra}{\rangle}
\nc{\la}{\langle}
\nc{\goto}{\rightarrow}
\nc{\swap}{\leftrightarrow}

\nc{\EE}[1]{ \mbox{$\cdot10^{#1}$} }
\nc{\abs}[1]{\left|#1\right|}
\nc{\at}[2]{\left.#1\right|_{#2}}
\nc{\norm}[1]{\|#1\|}
\nc{\abscut}[2]{\abs{#1}_{\scriptscriptstyle#2}}
\nc{\vek}[1]{{\rm\bf #1}}
\nc{\integral}[2]{\int\limits_{#1}^{#2}}
\nc{\inv}[1]{\frac{1}{#1}}
\nc{\dd}[2]{{{\partial #1}\over{\partial #2}}}
\nc{\ddd}[2]{{{{\partial}^2 #1}\over{\partial {#2}^2}}}
\nc{\dddd}[3]{{{{\partial}^2 #1}\over
        {\partial #2 \partial #3}}}
\nc{\dder}[2]{{{d #1}\over{d #2}}}
\nc{\ddder}[2]{{{d^2 #1}\over{d {#2}^2}}}
\nc{\dddder}[3]{{d^2 #1}\over
        {d #2 d #3}}
\nc{\dx}[1]{d\,^{#1}x}
\nc{\dy}[1]{d\,^{#1}y}
\nc{\dz}[1]{d\,^{#1}z}
\nc{\dl}[1]{\frac{d\,^{#1}l}{(2\pi)^{#1}}}
\nc{\dk}[1]{\frac{d\,^{#1}k}{(2\pi)^{#1}}}
\nc{\dq}[1]{\frac{d\,^{#1}q}{(2\pi)^{#1}}}
\nc{\dbar}{d\!\!\!\stackrel{\stackrel{\!-}{}}{}\!\!\!}
\nc{\cc}{\mbox{$c.c.$ }}
\nc{\hc}{\mbox{$h.c.$ }}
\nc{\cf}{cf.\ }
\nc{\erfc}{{\rm erfc}}
\nc{\Tr}{{\rm Tr\,}}
\nc{\tr}{{\rm tr\,}}
\nc{\pol}{{\rm pol}}
\nc{\sign}{{\rm sign}}
\nc{\bfT}{{\bf T }}

\nc{\cA}{{\cal A}}
\nc{\cB}{{\cal B}}
\nc{\cD}{{\cal D}}
\nc{\cE}{{\cal E}}
\nc{\cF}{{\cal F}}
\nc{\cG}{{\cal G}}
\nc{\cH}{{\cal H}}
\nc{\cL}{{\cal L}}
\nc{\cM}{{\cal M}}
\nc{\cO}{{\cal O}}
\nc{\cT}{{\cal T}}
\nc{\rvac}[1]{|{\cal O}#1\rangle}
\nc{\lvac}[1]{\langle{\cal O}#1|}
\nc{\rvacb}[1]{|{\cal O}_\beta #1\rangle}
\nc{\lvacb}[1]{\langle{\cal O}_\beta #1 |}
\nc{\bb}{\bar{\beta}}
\nc{\ctH}{\tilde{\cal H}}
\nc{\chH}{\hat{\cal H}}
\nc{\1}{\aa}
\nc{\2}{\"{a}}
\nc{\3}{\"{o}}
\nc{\4}{\AA}
\nc{\5}{\"{A}}
\nc{\6}{\"{O}}
\nc{\al}{\alpha}
\nc{\Del}{\Delta}
\nc{\e}{\epsilon}
\nc{\eps}{\epsilon}
\nc{\lam}{\lambda}
\nc{\om}{\omega}
\nc{\Om}{\Omega}
\nc{\ve}{\varepsilon}
\nc{\mn}{{\mu\nu}}
\nc{\k}{\kappa}
\nc{\vp}{\varphi}

\nc{\pub}[4]{\Bibitem{#1}#2, {\sl ``#3''}, #4.}
\nc{\advp}[3]{{\it  Adv.\ in\ Phys.\ }{{\bf #1} {(#2)} {#3}}}
\nc{\annp}[3]{{\it  Ann.\ Phys.\ (N.Y.)\ }{{\bf #1} {(#2)} {#3}}}
\nc{\apl}[3]{{\it  Appl. Phys. Lett. }{{\bf #1} {(#2)} {#3}}}
\nc{\apj}[3]{{\it  Ap.\ J.\ }{{\bf #1} {(#2)} {#3}}}
\nc{\apjl}[3]{{\it  Ap.\ J.\ Lett.\ }{{\bf #1} {(#2)} {#3}}}
\nc{\app}[3]{{\it Astropart.\ Phys.\ }{{\bf #1} {(#2)} {#3}}}
\nc{\cmp}[3]{{\it  Comm.\ Math.\ Phys.\ }{{ \bf #1} {(#2)} {#3}}}
\nc{\cqg}[3]{{\it  Class.\ Quant.\ Grav.\ }{{\bf #1} {(#2)} {#3}}}
\nc{\epl}[3]{{\it  Europhys.\ Lett.\ }{{\bf #1} {(#2)} {#3}}}
\nc{\ijmp}[3]{{\it Int.\ J.\ Mod.\ Phys.\ }{{\bf #1} {(#2)} {#3}}}
\nc{\ijtp}[3]{{\it Int.\ J.\ Theor.\ Phys.\ }{{\bf #1} {(#2)} {#3}}}
\nc{\jmp}[3]{{\it  J.\ Math.\ Phys.\ }{{ \bf #1} {(#2)} {#3}}}
\nc{\jpa}[3]{{\it  J.\ Phys.\ A\ }{{\bf #1} {(#2)} {#3}}}
\nc{\jpc}[3]{{\it  J.\ Phys.\ C\ }{{\bf #1} {(#2)} {#3}}}
\nc{\jpg}[3]{{\it J.~Phys.~G:~Nucl.~Part.~Phys.~}{{\bf #1} {(#2)} {#3}}}
\nc{\jap}[3]{{\it J.\ Appl.\ Phys.\ }{{\bf #1} {(#2)} {#3}}}
\nc{\jpsj}[3]{{\it J.\ Phys.\ Soc.\ Japan\ }{{\bf #1} {(#2)} {#3}}}
\nc{\lmp}[3]{{\it Lett.\ Math.\ Phys.\ }{{\bf #1} {(#2)} {#3}}}
\nc{\lncim}[3]{{\it  Lett.\ Nuov.\ Cim.\ }{{\bf #1} {(#2)} {#3}}}
\nc{\mpl}[3]{{\it  Mod.\ Phys.\ Lett.\ }{{\bf #1} {(#2)} {#3}}}
\nc{\ncim}[3]{{\it  Nuov.\ Cim.\ }{{\bf #1} {(#2)} {#3}}}
\nc{\np}[3]{{\it  Nucl.\ Phys.\ }{{\bf #1} {(#2)} {#3}}}
\nc{\pr}[3]{{\it Phys.\ Rev.\ }{{\bf #1} {(#2)} {#3}}}
\nc{\pra}[3]{{\it  Phys.\ Rev.\ }{{\bf A#1} {(#2)} {#3}}}
\nc{\prb}[3]{{\it  Phys.\ Rev.\ }{{\bf B#1} {(#2)} {#3}}}
\nc{\prc}[3]{{\it  Phys.\ Rev.\ }{{\bf C#1} {(#2)} {#3}}}
\nc{\prd}[3]{{\it  Phys.\ Rev.\ }{{\bf D#1} {(#2)} {#3}}}
\nc{\prl}[3]{{\it Phys.\ Rev.\ Lett.\ }{{\bf #1} {(#2)} {#3}}}
\nc{\pl}[3]{{\it  Phys.\ Lett.\ }{{\bf #1} {(#2)} {#3}}}
\nc{\prep}[3]{{\it Phys\. Rep.\ }{{\bf #1} {(#2)} {#3}}}
\nc{\prsl}[3]{{\it Proc.\ R.\ Soc.\ London\ }{{\bf #1} {(#2)} {#3}}}
\nc{\ptp}[3]{{\it  Prog.\ Theor.\ Phys.\ }{{\bf #1} {(#2)} {#3}}}
\nc{\ptps}[3]{{\it  Prog\ Theor.\ Phys.\ suppl.\ }{{\bf #1} {(#2)} {#3}}}
\nc{\physa}[3]{{\it  Physica\ A\ }{{\bf #1} {(#2)} {#3}}}
\nc{\physb}[3]{{\it  Physica\ B\ }{{\bf #1} {(#2)} {#3}}}
\nc{\phys}[3]{{\it Physica\ }{{\bf #1} {(#2)} {#3}}}
\nc{\rmp}[3]{{\it  Rev.\ Mod.\ Phys.\ }{{\bf #1} {(#2)} {#3}}}
\nc{\rpp}[3]{{\it Rep.\ Prog.\ Phys.\ }{{\bf #1} {(#2)} {#3}}}
\nc{\sjnp}[3]{{\it Sov.\ J.\ Nucl.\ Phys.\ }{{\bf #1} {(#2)} {#3}}}
\nc{\spjetp}[3]{{\it Sov.\ Phys.\ JETP\ }{{\bf #1} {(#2)} {#3}}}
\nc{\yf}[3]{{\it Yad.\ Fiz.\ }{{\bf #1} {(#2)} {#3}}}
\nc{\zetp}[3]{{\it Zh.\ Eksp.\ Teor.\ Fiz.\ }{{\bf #1} {(#2)} {#3}}}
\nc{\zp}[3]{{\it Z.\ Phys.\ }{{\bf #1} {(#2)} {#3}}}
\nc{\zpc}[3]{{\it Z.\ Phys.\ C\ }{{\bf #1} {(#2)} {#3}}}
\nc{\ibid}[3]{{\sl ibid.\ }{{\bf #1} {#2} {#3}}}
\newcommand{\minus}{\!-\!}
\newcommand{\plus}{\!+\!}
\nc{\rf}[1]{(\ref{#1})}
\nc{\nn}{\nonumber \\*}
\nc{\Lbmeff}{\cL^{\beta,\mu}_{\rm eff}}
\nc{\Fmn}{F_{\mu\nu}}
\nc{\Psibar}{\overline{\Psi}}
\nc{\ati}{\tilde{a}}
\nc{\bt}{\tilde{b}}
\nc{\kt}{\tilde{k}}
\nc{\mt}{\tilde{m}}
\nc{\pti}{\tilde{\mbox{\boldmath $\Pi$}}}
\nc{\pv}{\mbox{\boldmath $\Pi$}}
\nc{\para}{\parallel}
\nc{\orto}{\perp}
\nc{\LL}{Landau level}
\nc{\LLL}{lowest Landau level}
\nc{\amm}{anomalous magnetic moment}
\nc{\dega}{\Delta E^{\beta}_{\gamma}}
\nc{\dee}{\Delta E^{\beta,\mu}_{e^+e^-}}
\nc{\dele}{\Delta E}

\nc{\dmbm}{\Delta m^{\beta,\mu}}
\nc{\mhz}{{\hat{\mu}_z}}
\begin{document}
\baselineskip 20pt
%
%%%%%%%%%%%%%%%%%%%%%%%%%%%%%%%%%%%%%%%%%%%%%%%%%%%%%%%%%%%%
% TITLEPAGE
%%%%%%%%%%%%%%%%%%%%%%%%%%%%%%%%%%%%%%%%%%%%%%%%%%%%%%%%%%%%
%\large
\thispagestyle{empty}
%\begin{flushright}
\begin{flushright} CERN-TH/95-243\\
            G\"{o}teborg ITP 95-15 \\
hep-ph/9509418\\
September 1995\\
 \end{flushright}
\begin{center}
{\Huge\bf   Thermal Fermionic Dispersion \\[5mm] Relations
  in a Magnetic Field}\\[5mm]
\normalsize
\end{center}
%
%\vspace*{1mm}
\bc
{\large Per Elmfors,\footnote{\noindent Email address:
elmfors@cern.ch}}\raisebox{1ex}{,a}
{\large David Persson\footnote{\noindent Email address:
tfedp@fy.chalmers.se}}\raisebox{1ex}{,b}  and
{\large Bo-Sture Skagerstam\footnote{\noindent Email address:
tfebss@fy.chalmers.se}}\raisebox{1ex}{,a,b,c}
 \\[4mm]
{\sl \raisebox{1ex}{a}CERN, TH-Division,
CH-1211 Geneva 23, Switzerland \\ }
%\vspace*{1mm}
{\sl \raisebox{1ex}{b}Institute of Theoretical Physics,
   Chalmers University of Technology and \\
    University of G\"oteborg, S-412 96 G\"oteborg,
Sweden \\ }
{\sl \raisebox{1ex}{c}Department of Physics,
   University of Trondheim,\\
   N-7055 Dragvoll, Norway\\ }
\ec

%\vspace*{1mm}
%
\vfill
\bc
{\bf Abstract} \\
\ec
{\small
\begin{quotation}
\noindent
The thermal  self-energy of an electron in a static uniform
magnetic field $B$ is calculated to first order in the fine structure
constant $\alpha $ and to all orders in $eB$. We use two methods,
one based on the Furry picture and another
based on Schwinger's proper-time
method. As  external states we
consider relativistic Landau levels with special
emphasis on the  lowest
Landau level. In the high-temperature limit we derive self-consistent
dispersion relations for particle and hole excitations,
showing the chiral asymmetry caused by the
external field. For weak fields, earlier results
on the ground-state energy and the anomalous magnetic moment are
discussed and compared with the present analysis.
In the strong-field limit the appearance of a field-independent
imaginary part of the self-energy, related to Landau damping
in the $e^{+}e^{-}$ plasma, is pointed out.
\end{quotation}}
\vfill
\newpage
%%
%% -----------------------------------------------------------------
\normalsize
\setcounter{page}{1}
\baselineskip 20pt
%
%%%%%%%%%%%%%%%%%%%%%%%%%%%%%%%
\section{Introduction}
\label{intro}
%%%\Section{intro}{Introduction}
%%%%%%%%%%%%%%%%%%%%%%%%%%%%%%%
It has become increasingly important to understand
the quantum field theory
of elementary processes in the presence of both a thermal heat bath
and strong background fields, in particular in connection with
applications to astrophysical
and cosmological models. A natural strategy  for making progress in this
area is to generalize existing background field calculations to
finite temperature and vice versa.
The first obvious quantity to study is the effective potential,
which governs the thermodynamics of the system, and there
are several papers on this topic
\bort{\cite{CanutoC68,Dittrich79,ceo90,ElmforsPS94,%
ElmforsS95,ElmforsLPS95,BrandtF95}}
[1--7].
Next, we are interested in more detailed kinematical issues, which
are determined by the propagation of the elementary excitations in a plasma.
Dispersion relations for electrons have been studied in great detail
in strong background fields but without any heat bath
\bort{\cite{TsaiY73,Tsai74,SchwingerT74,BaierKS74,kobsak83,FerreObesoPR84%
,GepragsRHRW94}}
[8--14],
and at high temperature but in the absence of external fields
\bort{\cite{Klimov82,Weldon82,Weldon89,BraatenP90,FrenkelTW90}}
[15--19].

In this paper we perform a detailed study of the fermion self-energy in QED,
taking into account both the effects from a magnetic background field
and a thermal heat bath. In particular we consider the
limits of strong fields in the \LLL{}  and weak fields at high temperature.
There are two basic methods for doing perturbative quantum field
theory calculations
in a constant magnetic background field. One is the Furry picture, where the
fermion propagator is constructed by an explicit sum over the solutions
to the Dirac equation \cite{furry51}.
The other one is the Schwinger proper-time method,
where the propagator is expressed directly in terms of operators without
any reference to an explicit representation
         of the states \cite{Schwinger51}.
The different methods are suitable for different calculations and we have used
them both. Typically, the Furry picture is convenient in the strong field
limit where only a few \LL{}s contribute, while Schwinger's method is
particularly useful for
weak fields.

We  have organized the paper as follows. The two basic methods,
the Furry picture and Schwinger's proper-time method, are introduced
in Sections \ref{furry} and \ref{schwinger}. The weak field limit is studied
in Section \ref{weakb} using both methods. It is particularly interesting
to find the dispersion relation at high temperature for
weak fields, as
we do in Section \ref{ht},  since it is then consistent to not only calculate
first-order corrections but to solve the dispersion relation
 self-consistently.
As an application of the weak field expansion we study the \amm{}
in Section \ref{amm} and
compare it with other calculations. The strong field limit, where only
the \LLL{} contributes, is investigated in Section \ref{strongb}.
Discussions of the results follow in Section \ref{disc}, and the
appendices contain some technical details.
%
%%%%%%%%%%%%%%%%%%%%%%%%%%%%%%%%%%%%%%%%%%%
%%%\Section{furry}{Self-energy corrections of the lowest Landau level in the Furry picture}
\section{Self-energy corrections of the lowest Landau level in the Furry picture}
\label{furry}
%%%%%%%%%%%%%%%%%%%%%%%%%%%%%%%%%%%%%%%%%%
%
We consider Dirac fermions with charge $-e$
 in the presence of an external field
described by the vector potential $A_\mu=(0,0,Bx,0)$, corresponding to a
static uniform magnetic field in the negative $z$-direction.
 Using static energy
solutions to the Dirac equation
$(i\pa\sslask+eA\sslask-m)\Psi^{(\pm)}_\kappa=0$,
we may represent  the second quantized
fermion field in the Furry picture \cite{furry51} as
\be{qfield}
    \Psi=\sum_\kappa\left[b_\kappa\Psi^{(+)}_\kappa(\vek{x},t)+
    d^\dagger_\kappa\Psi^{(-)}_\kappa(\vek{x},t)\right]\ ,
\ee
where $\kappa$ denotes a complete set of quantum numbers and $(b,d)$ are
the standard annihilation operators for particles and antiparticles
\cite{ItzyksonZ80}.
The fermion  propagator, including the effects of some distribution of
particles, can then be constructed explicitly as the expectation value
(see \cite{ElmforsPS94} for more details)
\be{thpropfurry}
   iS(x',x)=\la \bf T \left[\Psi(x')\Psibar(x)\right] \ra~~~.
\ee
Evaluating the expectation values of the creation and annihilation operators
we may separate into  vacuum and thermal (actually more generally due to
some arbitrary distribution of fermions) contributions, $S(x',x)=
S_{\rm vac}(x',x)+S^{\beta,\mu}(x',x)$. We find
\bea{propform}
        iS_{\rm vac}(x',x)&=& \sum_\kappa \left[ \Theta(t'-t)
        \Psi^{(+)}_\kappa(x')\Psibar^{(+)}_\kappa(x) - \Theta(t-t')
        \Psi^{(-)}_\kappa(x')\Psibar^{(-)}_\kappa(x) \right]~~~, \\
        iS^{\beta,\mu}(x',x)&=& \sum_\kappa \left[ f_F^+(E_\kappa)\,
        \Psi^{(+)}_\kappa(x')\Psibar^{(+)}_\kappa(x) -  f_F^-(E_\kappa)
        \Psi^{(-)}_\kappa(x')\Psibar^{(-)}_\kappa(x) \right]~~~,
        \label{termprop}
\eea
where $E_\kappa$ is the energy eigenvalue of $\Psi^{(\pm)}_\kappa$.
For fermions, the antiparticle distribution is determined by
$ f^{-}_{F}(k_0)\equiv 1-f_{F}^{+}(-k_0)$. We shall here consider only
the case of thermal equilibrium:
\be{}
        f_F^{\pm}(k_0)= \frac{1}{e^{\beta(k_0 \mp \mu)}+1}\ ,
\ee
where $\beta$ is the inverse temperature and $\mu$ is the chemical
potential determined by the charge density of electrons and
positrons of the system.
For the explicit form of the wave-functions and of the propagator
 we have summarized the relevant expressions in Appendix \ref{app:efp}.

To lowest order in the electromagnetic coupling
there are two possible contributions to the electron self-energy,
the one-particle irreducible self-energy diagram and the tadpole.
In Appendix \ref{app:tadpole} we show that the tadpole gives no contribution
to the self-energy in a neutral background. Therefore,
in configuration space the self-energy is to one-loop order given by
\be{self-energy}
          -i\Sigma(x',x) = (-ie)^2\gamma_\mu iD^{\mu\nu}(x'-x)
        iS(x',x) \gamma_\nu~~~.
\ee
 In a general covariant gauge, parametrized by $\xi$, the photon propagator
is given by
\bort{\cite{KobesSW85,Landsman86,Kowalski87}}
[23--25]
\be{photprop}
         iD^{\mu\nu}(x)=\int [d^4q] e^{-iq\cdot x} \left( g^{\mu\nu}-
        \xi\, q^\mu q^\nu \frac{\partial}{\partial q^2}\right)  \left[
        \frac{-i}{q^2+i\ve} - 2\pi \delta(q^2)f_B(q_0) \right]~~~,
\ee
where the photon distribution function in the case of thermal equilibrium is
\be{}
        f_B(q_0)=\inv{\exp[\beta |q_0|]-1]}~~,
\ee
 and the operator $\pa/\pa q^2$
is not supposed to act on the $q_0$ in $f_B(q_0)$. We
have written $[d^n q] \equiv
d^n q/(2\pi)^n$ as a short-hand notation for the integration measure.
The expectation value, $\dele$, of the
self-energy  in a state described by the wave function
$\Psi^{(+)}_{\zeta;n,p_y,p_z}( x)$ is  defined by the expression
\be{expself}
        \dele_{\zeta;n,p_y}(p_z) = \frac{ \int d^4 x \,d^4x'\,
        \Psibar^{(+)}_{\zeta;n,p_y,p_z}( x') \Sigma(x',x)
        \Psi^{(+)}_{\zeta;n,p_y,p_z}( x)}
        {\int dy\,dz\,dt}~~~,
\ee
where $\int dy\,dz\,dt$ is a normalization factor from the continuous
spectrum in $p_y$, $p_z$, since $\int d^4 x \Psi^\dagger_\kappa(x)
        \Psi_\kappa(x)=\int dy\,dz\,dt$, cf. \eq{psinorm}.
In the gauge we use, the translational invariance in the plane perpendicular
to the $B$ field is reflected by the energy
degeneracy for different values of $p_y$.
Explicit
calculations show indeed that $\dele_{\zeta;n,p_y}(p_z)$ is
independent of $p_y$ in the \LLL,
 so we consider  for simplicity
only $p_y=0$ here. We formally show in Appendix
\ref{app:gauge} that the self-energy is independent of the gauge-fixing
parameter $\xi$ on the tree-level mass shell. This is explicitly verified
in the high-field limit in Section \ref{stronggauge}. We shall therefore use
the Feynman gauge $\xi=0$ for the photon propagator.
 From the inverse propagator to one-loop order,
 the effective Dirac equation is obtained as
\be{effdir}
        (i\not\!\!D - m)\Psi ({\bf x},t) = \int d^4x' \,
        \Sigma(x,x') \Psi(x')~~~,
\ee
where $D_\mu=\pa_\mu-ieA_\mu$.
Let us consider
 the value of the energy $E$, appearing in the wave functions for
particles in \eq{psiform} as $\exp[iEt]$, as
a free parameter not to be constrained by the
tree-level Dirac equation.
Then the expectation value of \eq{effdir} leads to the dispersion relation
\be{eshift}
        E=E_n(p_z)+  \dele(p_z)~~~,
\ee
where $E_n(p_z)=\sqrt{m^2+p_z^2+2eBn}$, and $\dele$ is to be calculated on
the tree-level mass shell.
It is tempting to try to solve \eq{eshift} self-consistently in $E$, by
substituting $E_n \rightarrow E$ in the phase factor in $\Psi$
before taking the expectation value,
and calculate $\dele$ with
such an arbitrary $E$. However, this is inconsistent for several reasons.
First, not only $E$ but
the actual form of the wave functions should then also be solved
self-consistently. Secondly,
it turns out that the self-energy is in general only gauge-invariant
on the tree-level mass shell and a self-consistent solution would
become gauge-dependent.
 In \cite{Mak94} a fermion propagator equivalent to the one
used here, but written explicitly in terms of $\gamma$ matrices, was
derived. Using this propagator and the general ansatz for the wave functions in
Eqs.~(\ref{psiform}) and (\ref{phiform}), it is possible to calculate
the general matrix structure of the self-energy operator, that acts on
the space-independent spinor $u_\kappa$,  in the
Furry picture.
Then one could proceed as in the following sections, where instead a
generalization of Schwinger's proper-time method has been employed,
and solve the self-consistent dispersion relation in a limit,
such as the high-temperature limit, where
 the off-shell gauge
dependence may be neglected.

Here we shall confine ourselves to the calculation
of the energy shift for
 an electron in the lowest relativistic Landau level.
Suppressing the $\zeta=1,\ n=0,\ p_y=0$ subscripts and the $B$ and $p_z$
dependence, we separate the self-energy into its
contributions from the vacuum, thermal
fermions, and thermal photons
\be{}
        \dele=\dele_{\rm vac} +\dee +\dega~~~,
\ee
respectively.
In \cite{kobsak83} the vacuum part was calculated for $p_z=0$,
 using the methods adop\-ted here.
Introducing an arbitrary momentum parallel to the external field does not
alter anything in the vacuum sector, since the relativistic invariance is
unbroken in the $z$ and $t$ directions. The self-energy is then a function of
 only $E_0^2-p_z^2=m^2$ and $eB$. We have confirmed this by explicit
calculations, of which we only quote the result here,
\bea{LLLvac}
        \dele_{\rm vac}&=& \frac{m^2}{E_0} \frac{\alpha}{2\pi} \int_0^1 ds\,
        \int_0^\infty du\,\exp(-m^2 u s^2)  \non \\
        &&\times\left\{ \frac{2eB \left[ 1+ s\,e^{-2eBus}\right]}
        {1+2eBu(1-s)-e^{-2eBus}}
        - \frac{1+s}u \right\}~~~.
\eea
The second term on the right-hand side comes from
 the ordinary mass renormalization to make
the vacuum part of the  self-energy vanish for $B=0$.

Using the thermal electron propagator given in Appendix \ref{app:efp},
we have
\bea{LLLthe}
        \dee\arreq-i\frac{2 e^2}{E_0}
        \frac{\int d^4x\,d^4x'}{\int dy'\,dz'\,dt'}
        \exp[iE_0(t-t')-ip_z(z'-z)] I_{0;0}(x)I_{0;0}(x') \non \\
        \align\times  \int [d^4 q]
        \frac{\exp[-i q \cdot (x-x')]}{q^2+i\ve} \nn
        \align\times\sum_{n=0}^\infty
        \int_{-\infty}^\infty  [d k_0]\,  [dk_y] \,[dk_z]
        \exp[-ik_0 (t'-t)+ik_y(y'-y)+ik_z(z'-z)] \non \\
        \align \times 2\pi i \delta(k_0^2-m^2-k_z^2-2eBn) \, f_F(k_0)\non \\
        \align \times\left\{m^2 I_{n;k_y}(x) I_{n;k_y}(x')+
        \left[ m^2-E_0 k_0+p_z k_z  \right]
                 I_{n-1;k_y}(x) I_{n-1;k_y}(x') \right\}~~~.
        \nn[-2mm]
\eea
The $t,y$ and $z$ integrals are trivially performed, and produce
$\delta$-functions that are used to perform the corresponding $q$
integrals.
We now express $I_{n;k_y}$ in terms of the  explicit form in \eq{Indef},
and use the fact that
$H_0(x)=1$. Then the variables $x,x',k_x \equiv q_x$ and $k_y$
 are shifted and rescaled. Fourier-transforming the $\delta$-function
\be{deltafour}
        \delta(k_0^2-m^2-k_z^2-2eBn)=\frac1{2\pi} \int_{-\infty}^\infty ds
        \exp[-is(k_0^2-m^2-k_z^2-2eBn) ]~~~,
\ee
and using the Feynman prescription to write
\be{oneover}
        \inv{a+i\epsilon}=-i\int_0^\infty du\, e^{iu(a+i\epsilon)}~~,
\ee
with $a=(k_0-E_0)^2-(k_z-p_z)^2-2eB(k_x^2+k_y^2)+i\ve$, we find
\bea{delcalc}
        \dee\arreq-i \frac{2e^2}{(2\pi)^5} \frac{2eB}{E_0}
        \int_{-\infty}^\infty  dk_0\,f_F(k_0)\int_{-\infty}^\infty ds
         \int_{0}^\infty du\, \int dx\,dx'\,
        dk_x\,dk_y\, dk_z         \non \\
        \align\times\exp\biggl\{-is[k_0^2-m^2-k_z^2]
        +iu[(k_0-E_0)^2-(k_z-p_z)^2-2eB(k_x^2+k_y)^2+i\ve]\biggr\} \non \\
        \align\times\exp\left[ -\frac12(x^2+x'^2)-k_y(x+x')+
        ik_x(x-x')-2k_y^2\right]
                         \non \\
        \align\times\left[m^2+( m^2-E_0 k_0+p_z k_z)\lambda \right]
        \sum_{n=0}^{\infty}\, \frac{\lambda^n}{n!}H_n(x)H_n(x')~~~,
\eea
where we have shifted the summation in the $I_{n-1,n-1}$ term, and defined
$\lambda \equiv e^{i2eBs}$.
Let us now use the integral representation of Hermite polynomials
\cite{gradshteyn}:
\be{hermint}
        \exp[-x^2/2]\,H_n(x)=\frac1{\sqrt{2\pi}}
        \int_{-\infty}^\infty dv\, \exp[-v^2/2+ivx]
        (-iv)^n~~~,
\ee
and similarly for $H_n(x')$ in terms of an integral over $v'$.
We may then identify an exponential function
\be{idexp}
        \sum_{n=0}^\infty\frac{\lambda^n(-iv)^n (-iv')^n}{n!}=
        \exp[-vv'\lambda]~~~.
\ee
We now have some Gaussian integrals that may be performed in the order
$v,x,v'$,
and finally $x'$. Let us  introduce a factor of convergence $\exp[-
        2eB\, \epsilon\,(k_x^2+k_y^2)]$, where $\epsilon \rightarrow 0^+$.
 We may now also change the order and perform the $k_x,k_y$
integrals,  while keeping the $s,u$ parameter integrals. The final result then
reads
\bea{dee}
        \dee \arreq
        -i\frac{\alpha }{\pi^2}\frac{eB}{E_0}
        \int_{-\infty}^\infty
        dk_0\,f_F(k_0)\int_{-\infty}^\infty ds \int_0^\infty\, du\,\int dk_z
        \exp\{-is[k_0^2-m^2-k_z^2] \} \non \\[2mm]
        \align\times\exp\bigl\{iu[(k_0-E_0)^2-(k_z-p_z)^2+i\ve]\bigr\}
        \frac{m^2+( m^2-E_0k_0+p_z k_z  )e^{i2eBs}}{1+2eB(\epsilon+iu)-
        e^{i2eBs}}~~~.\nn[-2mm]
\eea
The Gaussian integral over $k_z$ could easily be performed, but we prefer to
keep it in order to make the $s,u$ integrals simpler when considering the
weak field limit.

The thermal photon contribution is obtained in a way analogous
to the case of thermal electrons. The final result reads
\bea{dega}
        \dega&=&
        i\frac{\alpha }{\pi^2}\frac{eB}{E_0}
        \int_{-\infty}^\infty dk_0\,f_B(k_0)
        \int_{-\infty}^\infty ds \int_0^\infty\, du\,\int dk_z
        \exp\{-is[k_0^2-k_z^2] \} \non \\
        && \times\exp\{iu[(k_0+E_0)^2-m^2-(k_z+p_z)^2+i\ve]\}
        \frac{m^2-( E_0k_0-p_z k_z )e^{-i2eBu}}{1+2eB(\epsilon-is)-
        e^{-i2eBu}}~~~.\non \\
        &&
\eea
The two equations (\ref{dee}) and (\ref{dega}) serve as starting points for
the weak  field expansion in Section \ref{weakb}.
%
% PE 95.05.05
%%%%%%%%%%%%%%%%%%%%%%%%%%%%%%%%%%%%%%%%%%%%%%%%%%%%%%%%%%%%%
%%%\Section{schwinger}{Schwinger's proper-time method}
\section{Schwinger's proper-time method}
\label{schwinger}
%%%%%%%%%%%%%%%%%%%%%%%%%%%%%%%%%%%%%%%%%%%%%%%%%%%%%%%%%%%%%
%
In 1951 Schwinger \cite{Schwinger51}
used the analogy with a quantum mechanical
problem to find an explicit expression for the electron
propagator in an external electromagnetic field without
first finding the solutions to the Dirac equation in the
field. Later, mostly in the 70's, this method was used
to calculate a number of different properties  of electrons
in external fields.
The real-time finite-temperature propagator can be constructed in a simple
way from the zero-temperature propagator. Therefore, we shall
in this section  use
Schwinger's method to calculate the electron self-energy
in a magnetic field at finite temperature, in order to compare with the
corresponding calculations using the Furry picture, and to generalize to the
full self-energy operator. We use
standard real-time rules and as long as we are only
interested in the real part of the self-energy to one-loop it
is sufficient to calculate the (11)-part of the propagator.
The fact that standard Thermo Field Dynamics works also in a magnetic field
should be obvious, and it was discussed in \cite{ElmforsPS94,ElmforsS95}.
(For the imaginary part we can use the rules in
\cite{KobesS85}.) The propagator then is
\be{xprop}
        iS(x',x'')=\la x' |\frac{i}{\Pi\slask-m+i\ve}
        -f_F(p_0) \left(\frac{i}{\Pi\slask-m+i\ve}
        -\frac{i}{\Pi\slask-m-i\ve}\right)
        |x''\ra\ ,
\ee
where $f_F(p_0)$ is defined in \eq{fFdef}
and $\Pi\slask=\gamma^\mu(p_\mu+eA_\mu)$,as usual.
We shall not repeat Schwinger's calculation, but just give
the result for the zero temperature part in our notation
\bea{zprop}
        \la x'| \frac{i}{\Pi\slask-m+i\ve} |x''\ra &=&
        \frac{-i}{(4\pi)^2}\phi(x',x'')
        \int_0^\infty \frac{ds}{s} e^{-L(s)}
        \exp\left[-is\left(-\frac{e\sigma  F }{2}
        +m^2-i\ve\right)\right]
        \nn
        && \times\exp[-\frac{i}{4}(x'-x'') eF \coth eFs (x'-x'')]\nn
        &&\times
        \left\{\frac{\gamma}{2}(eF\coth eFs+eF)(x'-x'')+m\right\}
        \nn
        &\equiv &\phi(x',x'') \int [d^4p] e^{-ip(x'-x'')}i S_{\rm vac}(p)\ ,
\eea
where we have suppressed Lorentz indices ($F=F_\mu^{~\nu}$)
and used the notation
\bea{phi}
        \phi(x',x'')&=&\exp\left[ie\int_{x''}^{x'} dx^\mu\left(A_\mu
        +\inv{2}F_{\mu\nu}(x-x'')^\nu\right)\right]\ ,\nn
        \exp[-L(s)] &=& \frac{e^2s^2 a b}{\sin(eas)\sinh(ebs)}\ ,\nn[2mm]
        a^2-b^2 &=& B^2-E^2 \ ,\nn
        ab &=& \vek{B}\cdot\vek{E}\ .
\eea
The phase factor $\phi(x',x'')$  is the only part
of \eq{zprop} that depends
explicitly on the gauge, and that part  factors out in
a natural way when we compute the self-energy.
Restricting the field to be purely magnetic
in  the negative $z$-direction we find
in momentum space ($\sigma_z\equiv\sigma_{xy}$)
\bea{Bprop}
        iS_{\rm vac}(p)&=& \int_0^\infty ds \frac{e^{ieBs\sigma_z}}{\cos eBs}
        \exp\left[is\left(p^2_\para-\frac{\tan eBs}{eBs}p^2_\orto-m^2
        +i\ve\right)\right]\nn
        &&\times\left\{\gamma p_\para-
        \frac{e^{-ieBs \sigma_z}}{\cos eBs}\gamma p_\orto
        +m\right\}\ ,
\eea
where for general four-vectors $a$ and $b$,
$a\cdot b_\para=a_0 b_0-a_z b_z$ and
$a\cdot b_\orto=a_x b_x+a_y b_y$.
In order to calculate the thermal part we choose a gauge
where $A_0=0$ and $\pa_0 A_\mu=0$. This is very natural
for a static magnetic field (for a further discussion
of the gauge dependence of the thermal
propagator in a background field, see \cite{ElmforsS95}).
The $\phi(x',x'')$ can then also be factorized
out in front of the thermal propagator.
At finite temperature we see from \eq{xprop} that
 \eq{Bprop} should be replaced by
\be{thprop}
        iS_{\rm vac}(p)-f_F(p_0)\biggl(iS_{\rm vac}(p)-
        iS_{\rm vac}^*(p)\biggr)\ .
\ee
The real combination that occurs in the thermal part
of \eq{thprop} is finally obtained
by extending the $s$-integral
in \eq{Bprop}
from $-\infty$ to $\infty$. In the integrand of \eq{Bprop}
there are poles and essential singularities on the real
$s$ axis. They have to be avoided by taking the integration
contour in the lower half-plane for positive $s$ (see
e.g. \cite{ElmforsPS94} for a discussion of this contour); to
get a real quantity for the thermal part
it has, therefore, to go in the lower
half-plane also for negative $s$.
These poles are similar to the ones
responsible for the non-perturbative pair
production in an external electric field \cite{Schwinger51}
or the de~Haas--van~Alphen oscillations in a degenerate electron gas
\cite{ElmforsPS94}. When we consider the weak field limit in
Section \ref{weakb} we do not encounter these poles
to leading order.

The self-energy may be represented in momentum space as
\be{se}
        -i\Sigma(x',x'') =
        -i\phi(x',x'')\int [d^4p] e^{-ip(x'-x'')} \Sigma(p)\ ,
\ee
where the only breaking of translational invariance is in the
phase factor $\phi(x',x'')$.
There are two contributions to the real
part of the thermal self-energy at one loop.
One from thermal photons
and one from thermal electrons. Let us start with the photon
contribution. The electron contribution is completely
analogous. The thermal part of the photon propagator may
be represented in the Feynman gauge by
\be{thphprop}
        D_\beta^{\mu\nu}(k)=
        -g^{\mu\nu} f_B(k_0)\left(\frac{i}{k^2+i\ve}-
        \frac{i}{k^2-i\ve}\right)=-g^{\mu\nu}f_B(k_0)
        \int_{-\infty}^\infty
        dt \exp[it k^2-\abs{t} \ve]\ ,
\ee
and the corresponding contribution to the self-energy is
\bea{thphse}
        \Sigma^\beta_\gamma(p)\arreq ie^2 \gamma^\mu
        \int [d^4k] f_B(k_0)\int_{-\infty}^\infty dt
        \exp[it k^2-\abs{t} \ve] \nn
        \align\times\int_0^\infty ds \exp\left[is\left( (p-k)^2_\para
        -\frac{\tan eBs}{eBs} (p-k)^2_\orto-m^2+i\ve\right)\right]\nn
        \align\times\frac{e^{i eBs \sigma_z}}{\cos eBs}
        \left\{\gamma (p-k)_\para-\frac{e^{-i eBs \sigma_z}}{\cos eBs}
        \gamma(p-k)_\orto+m\right\}\gamma_\mu\nn[3mm]
        \arreq \frac{ie^2}{(2\pi)^3}\int_{-\infty}^\infty
         [dk_0] f_B(k_0) dt\,ds\,
        \left(\frac{\pi}{i(s+t)}\right)^{1/2}
        \frac{eB\pi}{i(eBt+\tan eBs)}\nn
        \align\times\exp\left[i\left(tk_0^2+s(p_0-k_0)^2-\frac{st}{s+t}p_z^2
        -\frac{t\tan eBs}{eBt+\tan eBs}p^2_\orto-m^2 s\right)
        -s\ve-\abs{t}\ve\right]\nn
        \align\times\gamma^\mu\frac{e^{i eBs \sigma_z}}{\cos eBs}
        \left\{\gamma_0(p_0-k_0)-\frac{t}{s+t}\gamma_z p_z
        -\frac{e^{-i eBs \sigma_z}}{\cos eBs} \frac{eBt}{eBt+\tan eBs}
        \gamma p_\orto+m\right\}\gamma_\mu\ ,\nn
\eea
where we have followed \cite{Tsai74} very closely, the only
essential difference
being that the $k_0$-integral cannot be carried out explicitly.
The $\Sigma^\beta_\gamma$ above is a not an operator but
a complicated function of the parameter $p_\mu$.
In order to obtain the expectation value of the
energy shift in a given state one
would have to multiply it with explicit wave functions and
integrate over $p_\mu$. There is, however, a clever way to rewrite
it as an operator \cite{Tsai74}
(thus replacing $p_\mu$ with gauge-invariant
operators $\Pi_\mu$) which, when acting on the tree
level wave functions, has simple properties. Noticing that
$\phi(x',x'')$ only depends on $x'_\orto$ and $x''_\orto$, we write
\be{seop}
        \la x'|\hat{\Sigma}|x''\ra=\int[d p_\para] e^{-ip(x'-x'')_\para}
        \phi(x',x'')\int [d p_\orto] e^{ip(x'-x'')_\orto}
        \Sigma(p_\para,p_\orto) \ .
\ee
The key relations to be used are then
\bea{oprel}
        \phi(x',x'')\align\int [d p_\orto] e^{ip(x'-x'')_\orto}
        \exp\left[-i\frac{\tan v}{eB}p^2_\orto\right]\nn
        \align=
         \la x'|\exp\left[-i\frac{v}{eB}\Pi^2_\orto\right]
        |x''\ra\cos v\ ,\nn
        \phi(x',x'')\align\int [d p_\orto] e^{ip(x'-x'')_\orto}
        \exp\left[-i\frac{\tan v}{eB}p^2_\orto\right]\gamma p_\orto\nn
        \align=
         \la x'|\exp\left[-i\frac{v}{eB}(\Pi^2_\orto
        -eB\sigma_z)\right]  \Pi\slask_\orto
        |x''\ra \cos^2 v\ .
\eea
After performing the $\gamma$-matrix algebra we obtain
the final expression for the
self-energy operator
\bea{thphseop}
        \hat{\Sigma}^\beta_\gamma&=&\frac{e^2}{8\pi^2}
        \int_{-\infty}^{\infty} [d k_0] \int_{-\infty}^{\infty} dt
        \int_{0}^{\infty} ds  \left(\frac{\pi}{i(s+t)}\right)^{1/2}
        \frac{eB}{eBt+\tan eBs} f_B(k_0) \nn
        &&\times\exp\left[i\left(t k_0^2+s(p_0-k_0)^2-\frac{st}{s+t}p_z^2
        -\frac{v}{eB}\Pi^2_\orto-m^2s\right)\right] \nn
        &&\times\frac{\cos v}{\cos eBs}\left\{-2 e^{-ieBs\sigma_z}
        \left(\gamma_0(p_0-k_0)-\frac{t}{s+t}p_z\gamma_z\right)
        +4m\cos eBs \right.\nn
        &&\left.+\frac{2\cos v}{\cos eBs}\frac{eBt}{eBt+\tan eBs}
        e^{i v \sigma_z} \Pi\slask_\orto\right\} \ ,
\eea
where
\be{betadef}
        \tan v=\frac{eBt \tan eBs}{eBt+\tan eBs} \ .
\ee
We notice that $\Pi^2_\orto$ and $\Pi\slask_\orto$
do not commute  but
$\Pi\slask_\orto$ is actually multiplied by
a function of  only
$\Pi^2_\orto-eB\sigma_z$ with which it
does commute, so the ordering is not a problem.
A similar analysis for thermal electrons gives
\bea{thelseop}
        \hat{\Sigma}^\beta_{e^+e^-}&=&-\frac{e^2}{8\pi^2}
        \int_{-\infty}^{\infty} [d k_0]\int_{0}^{\infty} dt
        \int_{-\infty}^{\infty} ds  \left(\frac{\pi}{i(s+t)}\right)^{1/2}
        \frac{eB}{eBt+\tan eBs} f_F(k_0) \nn
        &&\times\exp\left[i\left(t (p_0-k_0)^2+sk_0^2-\frac{st}{s+t}p_z^2
        -\frac{v}{eB}\Pi^2_\orto-m^2s\right)\right] \nn
        &&\times\frac{\cos v}{\cos eBs}\left\{-2 e^{-ieBs\sigma_z}
        \left(\gamma_0 k_0-\frac{t}{s+t}p_z\gamma_z\right)
        +4m\cos eBs\right.\nn
        &&\left.+\frac{2\cos v}{\cos eBs}\frac{eBt}{eBt+\tan eBs}
        e^{i v \sigma_z} \Pi\slask_\orto\right\}\ .
\eea
The matrix elements of $\hat{\Sigma}^\beta_{e^+e^-}$ and
$\hat{\Sigma}^\beta_\gamma$ depend of course on the basis in which they are
calculated. A suitable basis that diagonalizes the eigenvalues
$\kappa\equiv(E,n,p_y,p_z)$ should, with  the chiral representation of
 $\gamma_\mu$, have the form (see e.g. Appendix
\ref{app:efp}):
\bea{Psi}
        \Psi_\kappa(x)&=&\exp[-i(Et-p_z z-p_y y)]V_{n,p_y}(x)
        u(E,n,p_y,p_z)\ ,\nn
        V_{n,p_y}(x)&=&{\rm diag}\biggl(I_{n,p_y}(x),I_{n-1,p_y}(x),
        I_{n,p_y}(x),I_{n-1,p_y}(x)\biggr)\ ,
\eea
where $I_{n,p_y}(x)$ is defined in Appendix \ref{app:efp}
and $u(E,n,p_y,p_z)$ is a Dirac spinor, independent of
$x_\mu$.
With this choice of wave function, the space-time integrals of (see
\eq{expself})
\be{expse}
        \la\Psi_\kappa|\hat{\Sigma}|\Psi_{\kappa'}\ra=
        \frac{\int d^4x\,d^4x'\overline{\Psi}_\kappa(x)
        \Sigma(x,x') \Psi_{\kappa'}(x')}{\int \,dy \,dz\,dt}
\ee
can be carried out directly and give a $\delta$-function
in $\kappa-\kappa'$. There then remains a $4\times 4$  matrix that
can be diagonalized with a suitable choice of the spinors
$u(E,n,p_y,p_z)$.
%
% PE 95.05.05
%%%%%%%%%%%%%%%%%%%%%%%%%%%%%%%%%%%%%%%%%%%%
%%%\Section{weakb}{Weak-field expansion}
\section{Weak-field expansion}
\label{weakb}
%%%%%%%%%%%%%%%%%%%%%%%%%%%%%%%%%%%%%%%%%%%%
\nc{\mga}{\delta m^{\beta}_{\gamma}}
\nc{\mee}{\delta m^{\beta,\mu}_{e^+e^-}}
%%%%%%%%%%%%%%%%%%%%%%%%%%%%%%%%%%%%%%%%%%%%%%%%
In many physical applications the magnetic field is strong
enough to be important,
but still weak enough for an expansion in $eB/m^2$ to be useful.
It should, however, be noted that there are some fundamental
difficulties with the limit of weak fields.
At any finite field strength the eigenfunctions are \LL{}s and
there is not a  single eigenfunctions that continuously
goes over to a plane wave when $B\goto 0$. The Fourier
coefficients of the function $I_{n,p_y}(x)$ in \eq{Indef} are given by
\be{FourierI}
        \int dx\, e^{ikx} I_{n,p_y}(x)= i^n \sqrt{\frac{2}{n!}}
        \left(\frac{\pi}{eB}\right)^{1/4}
        \exp\left[-\frac{(k-ip_y)^2}{2eB}-\frac{p_y^2}{2eB}\right]
        H_n\left(k\sqrt{\frac{2}{eB}}\right)\ ,
\ee
and they cannot be expanded in powers of $B$.
It is, therefore, not clear that there exists a continuous
limit as $B\goto 0$ in general.
In fact, we know that in a background of degenerate
electrons there are de~Haas--van~Alphen oscillations
that do not have a series expansion in $B$ \cite{ElmforsPS94}.
In standard
first-order perturbation theory it is enough to calculate
the expectation value of the
perturbation in the unperturbed states. It would therefore be
tempting to use plane waves as external states for weak fields.
But, as explained above, the exact states are not approximately
equal to plane waves and we have to use the Landau levels
as a basis even for weak fields.
\jump
%%%%%%%%%%%%%%%%%%%%%%%%%%%%%%%%%%%%%%%%%%%%%
\subsection{\sc The Furry picture}
\Label{weakfurry}
%%%%%%%%%%%%%%%%%%%%%%%%%%%%%%%%%%%%%%%%%%%%
In order to obtain the weak-field limit  we expand the denominator in \eq{dee}:
\be{weakele}
        \frac{-i 2eB}{1+2eB(\epsilon+iu)-e^{i2eBs}} =
        \frac1{s-u+i\epsilon} -i \frac{eBs^2}{(s-u+i\epsilon)^2}+
        {\cal O}(eB)^2~~~.
\ee
In the limit $B \rightarrow 0$ we may close the contour and use Cauchy's
theorem
to integrate over $s$. The simple pole from the first term in
\eq{weakele} gives a contribution for
$k_0^2-m^2-k_z^2 >0$. We may then perform also the $t$ and $k_z$ integrals,
with the final result
\bea{bzeroele}
\lefteqn{\dee(B=0) =\frac{\alpha}{2\pi} \frac{m}{E_0}
        \int_{-\infty}^\infty
        dk_0 \Theta(k_0^2-m^2)     f_F(k_0)}\non \\
        &&\times  \left[ 2 \frac{\sqrt{k_0^2-m^2}}m -
        \frac m{p_z}\, \ln \left(
        \frac{E_0 k_0 -m^2+p_z\sqrt{k_0^2-m^2}}{E_0 k_0
        -m^2-p_z\sqrt{k_0^2-m^2} }\right) \right]~~~.
\eea
This result is finite, well-behaved as $p_z \rightarrow 0$,
and it agrees with the thermal self-energy calculated in a conventional
plane-wave basis.
To linear order in $eB$ we use \eq{weakele} together with $e^{i2eBs}=
1+i2eBs+ {\cal O}(eB)^2$ in \eq{dee}.
In order not to get a contribution when closing the
$s$ contour by a semi-circle at infinity,
it is necessary to rewrite terms such as e.g. $s/(s-u+i\ve)=1+u/(s-u+i\ve)$
and treat the $1$ separately.
The term linear in $eB$ is thus obtained from
\be{linele}
        -i\, eB \left[ (E_0 k_0- p_z k_z ) + 2m^2\frac{ u}{s-u+i\epsilon} +
        (2m^2-E_0 k_0 +p_z k_z) \frac{u^2}{(s-u+i\epsilon)^2} \right]~~~.
\ee
 Integrating over $s$ the first term in \eq{linele} gives
a contribution proportional to
$\delta(k_0^2-m^2-k_z^2)$. For the last two terms we proceed as before,
close the $s$-contour and then use Cauchy's theorem. Performing the
standard integrals over $t$ and $k_z$ we arrive at the term linear in $eB$
\bea{blinele}
        \lefteqn{\dee(B)-\dee(0)\simeq \frac{\alpha}{4\pi} \frac{eB}{E_0}
        \int_{-\infty}^\infty dk_0 \,
        \frac{\Theta(k_0^2-m^2)}{\sqrt{k_0^2-m^2}} f_F(k_0) \frac1{p_z^2}}
                \non \\
        &&\times  \left[ \frac{\sqrt{k_0^2-m^2}}{p_z}(2E_0 k_0 -m^2)
        \ln \left(
        \frac{E_0 k_0 -m^2+ p_z\sqrt{k_0^2-m^2}}{E_0 k_0 -m^2-p_z
        \sqrt{k_0^2-m^2}} \right) \right. \non \\
        &&- 2 \left. \frac{(k_0^2-m^2)(2k_0^2-3E_0 k_0+m^2)+
         p_z^2(k_0^2+p_z^2-E_0 k_0)}{(k_0-E_0)^2} \right]~~~.
\eea
This expression has an ostensible singularity at $k_0=E_0$. When expanding
around this value we find that the possible divergences cancel, and that the
integral is finite as it stands for $p_z>0$.
The limit $p_z \rightarrow 0$ is considered in Section \ref{amm}.

The thermal photon contribution is obtained in a similar manner. We may in
this case also perform the $k_0$ integrals. In the field-free case we
obtain the well-known result
\bort{\cite{FujimotoHY82,PeressuttiS82,DonoghueHR8485,JohanssonPS86}}
[29--32]
\be{bzeroga}
        \dega(B=0)= \frac{m}{E_0}
        \, \alpha \frac\pi3 \frac{T^2}{m^2}~~~.
\ee
Similarly we find the term linear in $eB$
\be{blinga}
        \dega(B)-\dega(0)\simeq
        \frac{m^2}{E_0}\,  \frac{\alpha\pi}{6} \frac{T^2}{m^2} \frac{eB}{p_z^2}
        \left[ \frac{E_0}{p_z} \ln \left( \frac{ E_0+p_z}{ E_0-p_z} \right)
                -2 \right]~~~.
\ee
However, these results only apply to an external electron in the \LLL\,
its
energy being given according to the tree-level Dirac equation. In the next
subsection we shall
consider the general case.
%%%%%%%%%%%%%%%%%%%%%%%%%%%%%%%%%%%%%%%%%%%%
\subsection{\sc Schwinger's method}
\Label{weakschwinger}
%%%%%%%%%%%%%%%%%%%%%%%%%%%%%%%%%%%%%%%%%%%
In the weak-field limit the Landau levels get closer and closer,
and in the Furry picture an explicit method of resumming them
is always necessary. This resummation was performed in Section \ref{furry},
 and the result expanded to linear order in the external field in
the preceding
section.
However,  this problem does not occur when using  the Schwinger
proper-time method, since no explicit reference to the  Landau levels
is made. In fact, with Schwinger's method we have an expression in
terms of operators which we can apply to any state we wish. We shall
now use this method to study the weak-field limit of the self-energy.
Starting from
\eq{thphseop} we find ($\pv^2=\Pi_\orto^2+p_z^2$)
\bea{thphseopwb}
        \hat{\Sigma}^\beta_\gamma&=&-\frac{ie^2}{4\pi^{3/2}}
        \int_{-\infty}^\infty [d k_0]\int_{-\infty}^\infty dt
        \int_{0}^\infty ds f_B(k_0)\inv{(i(s+t))^{3/2}}
        \nn
        &&\times\exp\left[i\left(t k_0^2+s(p_0-k_0)^2-\frac{st}{s+t}
        \pv^2 -m^2s\right)\right] \nn
        &&\times\left\{e^{-ieBs\sigma_z}\left(\gamma_0(p_0-k_0)-
        \frac{t}{s+t}\gamma_z p_z\right)-2m
        -\frac{t}{s+t}e^{i\frac{st}{s+t}eB\sigma_z}
        \Pi\slask_\orto\right\} \ ,
\eea
after expanding most terms to linear order in $B$ (the terms
with poles as a function of $s$).
We keep the full $B$ dependence wherever it is added
linearly to $\pv^2$ since $\pv^2=p_z^2+eB(2n+1)$ and $n$ need not be
small. In all other places the dependence is $\cO(B^2)$.
The $t$ and $s$ integrals in \eq{thphseopwb} can be
performed using
\bea{tint}
        \int_{-\infty}^{\infty}\frac{dt}{(it)^{3/2}}
        \exp\left[i(ta+\frac{b}{t})\right] \arreq
        2\left(\frac{\pi}{b}\right)^{1/2}
        \sin\left(2\sqrt{ab}\right)\Theta(a)\ ,
        \nn
        \int_{-\infty}^{\infty}\frac{dt}{(it)^{3/2}}
        \inv{t}\exp\left[i(ta+\frac{b}{t})\right] \arreq
        2i\frac{(a\pi)^{1/2}}{b}\left(
        \frac{\sin\left(2\sqrt{ab}\right)}{2\sqrt{ab}}-
        \cos\left(2\sqrt{ab}\right)\right)
        \Theta(a)\, ,
        \\
        \int_{-\infty}^{\infty}\frac{dt}{(it)^{3/2}}
        t\exp\left[i(ta+\frac{b}{t})\right] \arreq
        -2i\left(\frac{\pi}{a}\right)^{1/2}
        \cos\left(2\sqrt{ab}\right)
        \Theta(a)\ ,\non
\eea
for $b>0$, and
\bea{sint}
        i\int_0^{\infty} ds e^{isa}\sin (bs)\inv{s}&=&
        \inv{2}\ln\abs{\frac{a-b}{a+b}}\ ,
         \nn
        i\int_0^{\infty} ds e^{isa}\sin (bs)i&=&
        \frac{b}{a^2-b^2} \ ,
        \nn
        i\int_0^{\infty} ds e^{isa}\left(
        \frac{\sin (bs)}{bs}-\cos (bs)\right)\frac{i}{s}&=&
        -1-\frac{a}{2b}\ln\abs{\frac{a-b}{a+b}}\ ,
         \\
        i\int_0^{\infty} ds e^{isa}\left(
        \frac{\sin (bs)}{bs}-\cos (bs)\right)\frac{i}{s}is&=&
        -\frac{1}{2b}\ln\abs{\frac{a-b}{a+b}}
        -\frac{a}{a^2-b^2}\ .\non
\eea
The final result for the photon contribution is ($k=\abs{k_0}$,
$\pv\slask=\gamma\Pi_\orto+\gamma_z p_z$)
\bea{finph}
        \hat{\Sigma}^\beta_\gamma=
        -\frac{e^2}{2\pi}\int_{-\infty}^\infty [dk_0] f_B(k_0)
        \align\left\{ \inv{2\pti}\ln\abs{\frac{\ati-\bt}{\ati+\bt}}
        [\gamma_0(p_0-k_0)-2m-\pv\slask\,]
        \right. \nn \align\left.
        ~~-\frac{k}{\pti^2}\left(1+\frac{\ati}{2\bt}
        \ln\abs{\frac{\ati-\bt}{\ati+\bt}}\right)
        \pv\slask\right\} \ ,
\eea
where ($\pti=\sqrt{\pti^2}$)
\be{abph}
        \left\{\ba{lll}
        \ati&=&p_0^2-\pti^2-\mt^2-2p_0k_0\ ,\nn[2mm]
        \bt&=& 2 \pti k\ .
        \ea\right.
\ee
The meaning of $\mt^2$ and $\pti^2$ depends on which
$\gamma$-matrix they multiply,
and they should be replaced according to
\bea{mtpt}
        \mt^2&=&\left\{\ba{ll} m^2~,& {\rm when\ multiplying\ \ }
                \id,\gamma_x,\gamma_y\ ,\\
                m^2+eB\sigma_z~, &{\rm when\ multiplying\ \ }
        \gamma_0,\gamma_z\ ,\ea
        \right.\\
        \pti^2&=&\left\{\ba{ll} \pv^2~,& {\rm when\ multiplying\ \ }
        \id,\gamma_0,\gamma_z\ ,\\
                \pv^2-eB\sigma_z~, &{\rm when\ multiplying\ \ }
        \gamma_x,\gamma_y\ .\ea
        \right.
\eea
These are just replacement rules that simplify the writing and not
any mathematical identities.
The Equation \ref{finph} equals exactly the self-energy in the absence of the
magnetic field, with a replacement $p_\mu \rightarrow \Pi_\mu$,
 except that  $m^2$ and $\pv^2$
should be shifted by $\pm eB\sigma_z$ according to the
rules in \eq{mtpt}. We write it in this short-hand way in
order to more easily see that it reduces to the well-known
result in the limit of vanishing $B$. At the same time we
notice that it is not enough to simply replace $p_\mu$ by $\Pi_\mu$,
but there are some extra $\pm eB\sigma_z$ terms that enter.
The electron contribution is calculated in a similar way:
\bea{finel}
        \hat{\Sigma}^\beta_{e^+e^-}&=&
        \frac{e^2}{2\pi}\int_{-\infty}^{\infty}
         [d k_0] f_F(k_0) \Theta(k_0^2-\mt^2)\nn
        &&
        \times\left\{\frac{1}{2\pti}\ln\abs{\frac{\ati-\bt}{\ati+\bt}}
        [\gamma_0 k_0 -2m ]
%       \right.\nn &&\left.
        +\frac{\kt}{\pti^2}\left(1+\frac{\ati}{2\bt}
        \ln\abs{\frac{\ati-\bt}{\ati+\bt}}\right)\pv\slask
        \right\}\ ,
\eea
where now $\kt=\sqrt{k_0^2-\mt^2}$ and
\be{abel}
        \left\{\ba{lll}
        \ati&=&p_0^2-\pti^2+\mt^2-2p_0k_0\ ,\nn[2mm]
        \bt&=& 2 \pti \kt\ .
        \ea\right.
\ee
Again, it is easy to check that the zero-field limit from \eq{finel}
agrees with standard results.

One advantage with \eqs{finph}{finel} is that they are expressed directly
in terms of gauge-invariant operators and they can be used to calculate
the expectation value between any \LL{}s. It is particularly useful in Section
\ref{ht} where we solve a self-consistent dispersion relation without
  specifying the exact form of the spinors. On the other hand, the results
agree with the ones from
the Furry picture where we have checked them. An example is the \amm{}
that we discuss in Section \ref{amm}.
%
% PE 95.05.05
%%%%%%%%%%%%%%%%%%%%%%%%%%%%%%%%%%%%%%%%%%%%%%%
%%%\Section{ht}{High-temperature limit}
\section{High-temperature limit}
\label{ht}
%%%%%%%%%%%%%%%%%%%%%%%%%%%%%%%%%%%%%%%%%%%%%%%%%%
Since a few years back there has been a great interest in the
high-temperature limit of gauge theories,
mainly stimulated by the successful resummation of
a consistent and infinite set of diagrams, which is encoded
in the so-called Hard Thermal Loop (HTL) \cite{FrenkelTW90,BraatenP90}
effective action.
In the high-temperature limit it is meaningful to not only
compute a perturbative correction to the energy spectrum,
but also to solve the dispersion relation self-consistently.
The reason is that in QED the dominating $\cO (eT)$ correction
comes only from the one-loop diagram that we have calculated,
all higher-order diagrams being suppressed by extra factors
of $\cO(e^2T)$. The leading terms are also gauge-fixing independent and
gauge-invariant, which makes the whole procedure consistent.
Therefore, it is  particularly
interesting to take  the high-temperature limit
($T\gg m,~p_z,~\sqrt{eB},~\mu$) and study
the effects of a weak magnetic field.
The effective Dirac equation is
\bea{Deq}
        &&[\Pi\slask-m-\hat{\Sigma}(p_0,p_z,\pv_\orto)]\Psi=\nn[3mm]
        &&\left[s(p_0,\pv^2)\gamma_0p_0 -
        r(p_0,\pv^2)\gamma_zp_z-
        r(p_0,\pv^2-eB\sigma_z)\Pi\slask_\orto
        -m\right]\Psi=0\ ,
\eea
where as before $\pv^2=\Pi_\orto^2+p_z^2$ and
\bea{sandr}
        s(p_0,\pv^2) &=& \left(1-\frac{\cM^2}{2p_0\abs{\pv}}
        \ln\abs{\frac{p_0+\abs{\pv}}{p_0-\abs{\pv}}}\right) \ ,
        \nn[2mm]
        r(p_0,\pv^2) &=&\left(1+\frac{\cM^2}{\pv^2}
        \left(1-\frac{p_0}{2\abs{\pv}}
        \ln\abs{\frac{p_0+\abs{\pv}}{p_0-\abs{\pv}}}\right)\right)\ .
\eea
The temperature dependence enters only through the thermal mass
 $\cM^2=e^2T^2/8$. It is almost possible to guess the expression
in \eq{Deq} from the standard expression for the HTL Dirac equation.
The usual momentum $p_\mu$ should be replaced with the gauge-invariant
momentum $\Pi_\mu$, but there is an ambiguity in replacing $p^2$ by $\Pi^2$
or by $\Pi\slask\,\,\Pi\slask\,\,$. The correct way follows from the lengthy
calculations in Section \ref{weakb}. It would also be interesting to compare
\eq{Deq} with the equation of motion obtained
directly from the HTL effective action. This is in general difficult
due to the non-local
and non-linear character of the HTL effective action, but we have
checked that it agrees up to linear order in $B$.

{}From now on we shall take the high-temperature limit and neglect the vacuum
mass, which simplifies the Dirac equation considerably.
The matrix structure of the function $r(p_0,\pv^2-eB\sigma_z)$ can be
made explicit by rewriting it as
\be{exprB}
        r(p_0,\pv^2-eB\sigma_z)=\inv{2}(\id+\sigma_z)
        r(p_0,\pv^2-eB)
        +\inv{2}(\id-\sigma_z)
        r(p_0,\pv^2+eB)\ .
\ee
In the chiral
representation \eq{chiralgam} we obtain the following Dirac equation
\be{4x4Deq}
        \left(\ba{cc}
        0 & 0 \\
        0 & 0 \\
         -  p_0 s +  p_z r & r_-\xi_+ \\
        r_+\xi_-
        &-  p_0 s -  p_z r  \ea
        \ba{cc}
          -  p_0 s -  p_z r  & -r_-\xi_+\\
         -r_+\xi_-
        & -  p_0 s +  p_z r \\
         0 & 0\\
         0 & 0 \ea
        \right) \Psi=0\ ,
\ee
where $\xi_\pm=\Pi_x\mp i\Pi_y$ and $r_\pm=r(p_0,\pv^2\pm eB)$.
 Using the ansatz
in \eq{Psi}, the relations in \eq{xirel},
and dividing $\Psi$ into left and right
2-component spinors ($u=(R,\, L)^T$ in the
conventions of \cite{ItzyksonZ80}), we find the decomposed Dirac equations
\bea{LRDeq}
        \left(\ba{cc} -Es_n-p_zr_n & -i \sqrt{2eBn} r_{n-\inv{2}} \\
        i \sqrt{2eBn} r_{n-\inv{2}} & -Es_{n-1}+p_zr_{n-1}
        \ea\right) L &=& 0\ ,\nn[3mm]
        \left(\ba{cc} -Es_n+p_zr_n & i \sqrt{2eBn} r_{n-\inv{2}} \\
        -i \sqrt{2eBn} r_{n-\inv{2}} & -Es_{n-1}-p_zr_{n-1}
        \ea\right) R &=& 0\ ,
\eea
where $s_{n}=s(E,p_z^2+eB(2n+1))$, and similarly
for $r_n$. The dispersion relations
follow immediately by taking the determinants of \eq{LRDeq}
\bea{LRdr}
        \align L:\quad (Es_n+p_zr_n)
        (Es_{n-1}-p_zr_{n-1})-2eBnr^2_{n-\inv{2}}=0\ , \nn[3mm]
        \align R:\quad (Es_n-p_zr_n)
        (Es_{n-1}+p_zr_{n-1})-2eBnr^2_{n-\inv{2}}=0\ .
\eea
These relations are valid for all $n\geq 1$, but in the \LLL{} there is only
one non-zero component for each of $L$ and $R$,
so the dispersion relations reduce
to
\bea{LLLdr}
        \align L:\quad Es(E,p_z^2+eB)+p_zr(E,p_z^2+eB)=0\ , \nn
        \align R:\quad Es(E,p_z^2+eB)-p_zr(E,p_z^2+eB)=0\ .
\eea
Since the high-temperature correction is a consistently resummed
large correction to the vacuum dispersion relations, we want to
solve them self-consistently.
In the absence of a magnetic field it is well known that the high-temperature
dispersion relation for fermions has two branches
\cite{Klimov82,Weldon82,Weldon89} and that spin up/down (or left/right
handedness) is degenerate. The hole branch corresponds to a positive energy
solution to the factor in the dispersion relation that usually only gives a
negative energy solution. We have a similar phenomenon here, and the hole
solution is associated with the factor $Es_n+p_zr_n$ for positive $p_z$, even
though the dispersion relation does not factorize
completely. There are almost the separate
symmetries ($p_z\leftrightarrow-p_z$),  ($E\leftrightarrow-E$)
and ($L\leftrightarrow R$), but they fail
because of the difference in the index $n$ for the two factors
in \eq{LRdr}, which also
breaks the spin degeneracy.
We  notice on the other hand that there {\it are}  symmetries under
the combinations
($L\leftrightarrow R$, $p_z\leftrightarrow-p_z$),
($L\leftrightarrow R, E\leftrightarrow -E$) and
($E\leftrightarrow -E, p_z\leftrightarrow -p_z$). The number of states for
given values of $p_z$ and $n$ is eight, corresponding to
$(L/R)\times$ (particle/hole) $\times$ (positive/negative energy),
apart from the usual degeneracy in $p_y$.
%
%\jump
%
%
\bfig[t]
   \epsfxsize=15cm
   \epsfbox{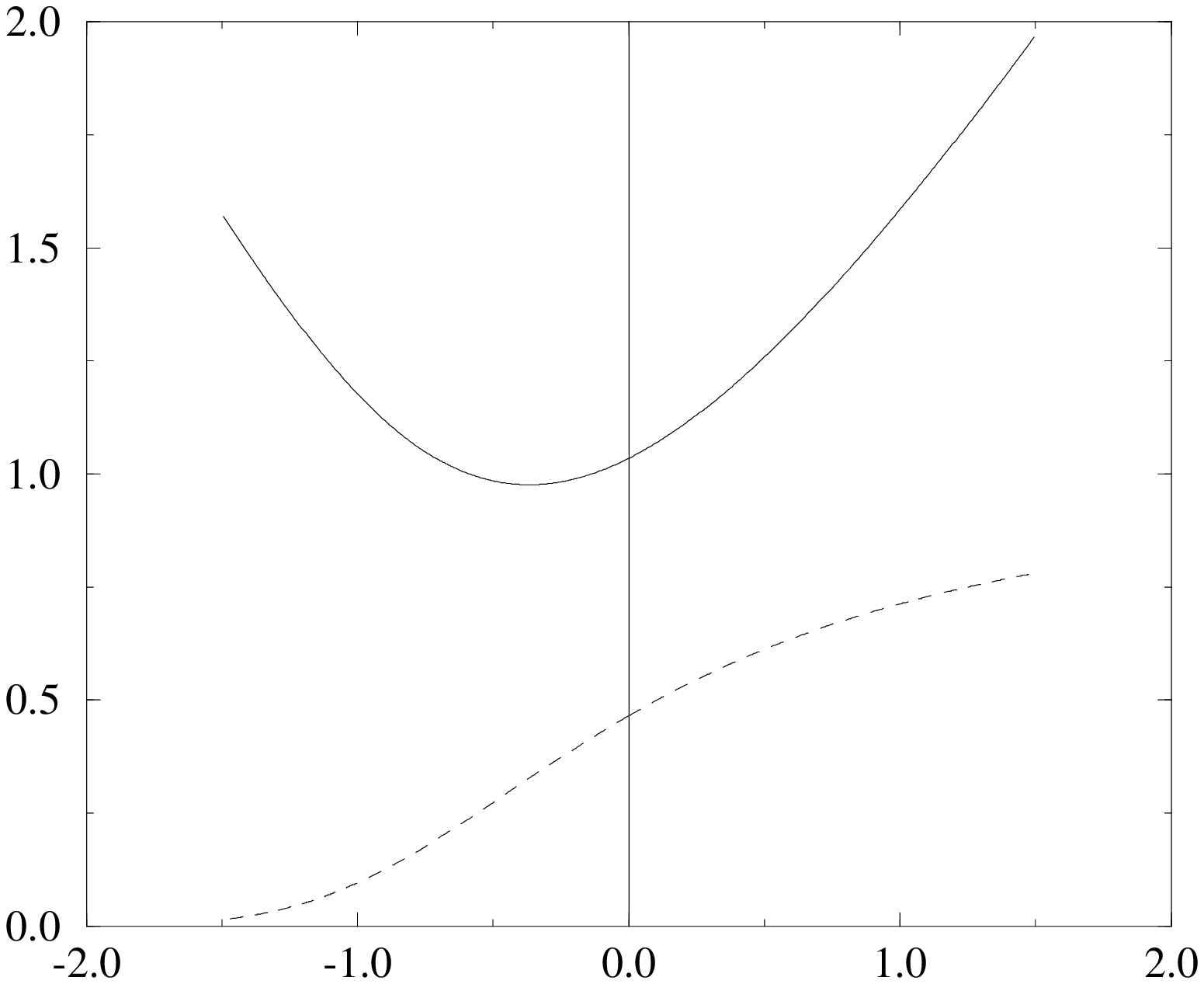}
   \setlength{\unitlength}{1mm}
\begin{picture}(0,0)(0,0)
   \put(0,0){}
   \put(225,40){\large $p_z$}
   \put(12,205){\large $E$}
   \put(175,115){\large $Z_h$}
   \put(270,165){\large $Z_p$}
   \put(175,220){\large $E_{Rh}$}
   \put(240,250){\large $E_{Rp}$}
   \put(160,300){\large $eB=0.2$}
   \put(160,285){\large $n=0$ \normalsize}
\end{picture}
%   \vspace{5mm}
   ~\\[-10mm]
   \figcap{Dispersion relation and spectral weight
   for the right-handed branch in the lowest Landau level ($n=0$).
   All dimensionful parameters are given in units of the thermal
   mass $\cM$.}
   \label{f:LLL}
\efig
\bfig[t]
   \epsfxsize=15cm
   \epsfbox{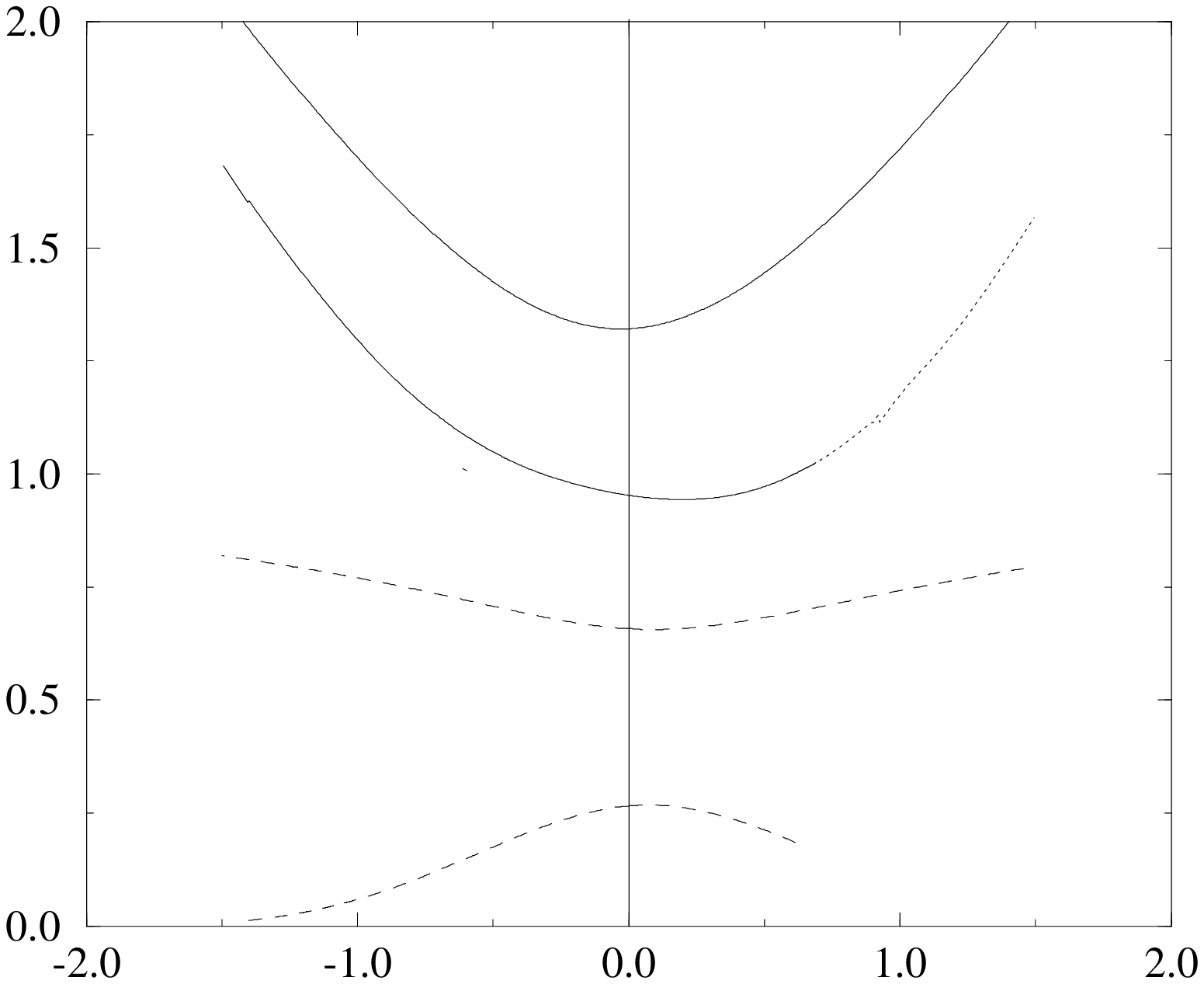}
   \setlength{\unitlength}{1mm}
\begin{picture}(0,0)(0,0)
   \put(0,0){}
   \put(225,40){\large $p_z$}
   \put(14,205){\large $E$}
   \put(240,115){\large $Z_h$}
   \put(240,169){\large $Z_p$}
   \put(240,206){\large $E_{Rh}$}
   \put(240,270){\large $E_{Rp}$}
   \put(170,310){\large $eB=0.2$}
   \put(170,295){\large $n=1$}
\end{picture}
%   \vspace{5mm}
   ~\\[-10mm]
   \figcap{Dispersion relations and spectral weights
   for the next-to-lowest Landau level ($n=1$).
   The dotted part of $\omega_h$ shows a continuation of the dispersion
   relation beyond the point where $\omega_h$ picks up an imaginary
   part.    All dimensionful parameters are given in units of the thermal
   mass $\cM$.}
   \label{f:NLLL}
\efig
In the \LLL{}, only right-handed particles and left-handed holes can propagate
for positive $p_z$, and vice versa for negative $p_z$ (see \eq{LLLdr}),
as can be understood from the following argument.
The energy shift for a magnetic moment  $\mbox{\boldmath $\mu$}$
is  given by
$-\mbox{\boldmath $\mu$}\cdot {\bf B}$. The electron-spin contribution to
 $\mbox{\boldmath $\mu$}$ points opposite to the spin,  since the
charge is negative.
 We have chosen $\vek{B}$
to point in the negative $z$-direction so that the energy is lowest when
the spin points in the positive $z$-direction. For
$p_z>0$ the helicity should thus be positive in the \LLL{} and we expect
the positive chirality state $R$ to propagate (see \fig{f:LLL}).
Left-handed particles have the wrong  helicity to be in the \LLL{}
for $p_z>0$. However, holes have opposite chirality--helicity relation, so
a left-handed hole can propagate for $p_z>0$. Similarly,
right-handed holes propagate for $p_z<0$ as shown in \fig{f:LLL}.
A similar asymmetry
exists in higher \LL{}s in the sense that, for a right-handed particle with
momentum $p_z$ and  $\pv^2=p_z^2+eB(2n+1)$,
there exists another right-handed particle
state with momentum $-p_z$ but with $\pv^2=p_z^2+eB(2n-1)$
and a different energy.
It is clear that this
asymmetry can be important for decay processes where only one chirality is
produced, for instance in $\beta$-decay. To get an asymmetry of this form, we
need a chiral charged particle with spin (or magnetic moment).
It clearly needs to
have a splitting between spin up and down, but it is also important that
it is  chiral in
order to separate right- and left-handed particles at the same time.
At zero temperature there
are no chiral charged spin $1/2$ particles in the Standard Model, but at
high temperature the dominant mass effect is the thermal one,
which is chirally invariant. In this way the
temperature effect makes the electron essentially
chiral without reducing its magnetic moment.

It has been observed \cite{Chugai84,DorofeevRT85} that the polarization
of electron and positron spins in a strong magnetic field generates
an anisotropy of the neutrino emission in weak processes, and that this
anisotropy could be the explanation of the high space velocities of pulsars.
In the presence of a thermal heat bath, we have seen above that a
further left--right asymmetry is generated for the electrons and positrons.
It remains to be seen to what extent this asymmetry increases
the asymmetry in the neutrino emission during the hot phase of the
supernova explosion.

In order to find out the correct final-state factors for a decay into the
different bran\-ches, not only the spectrum is needed but also the spectral
weight for each branch. That is, we need the wave function renormalization
factor $Z(k)$. It can be computed from \cite{Weldon89}
\be{Zdef}
        Z_i(p_z,n)^{-1}= \at{\frac{d}{d\om}}{\om=E_i(p_z,n)}
        \left(\Tr\left[(\cD(\om,p_z,n)\gamma_0)^{-1}\right]\right)^{-1}\ ,
\ee
where $\cD(\om,p_z,n)$
is the matrix in \eq{4x4Deq} and $E_i$ corresponds to the
solutions for $L$, $R$, particle and hole. The spectral functions
should satisfy the
sum rule for $n\geq 1$
\be{Zsum}
        Z_{Lp}+Z_{Lh}+Z_{Rp}+Z_{Rh}+{\rm multiparticle~states}=2\ .
\ee
We use a definition of $Z$ that differs from \cite{Weldon89} by a factor of
 2.
In the \LLL{} there are only half as many states, so the spectral weights
add up to 1.
The result of a numerical calculation
of the spectral weight as well as of the spectrum and for the \LLL{}
is presented in \fig{f:LLL}.
It shows the right-handed branch, but the picture is the same for the
left-handed
branch with $p_z\leftrightarrow -p_z$.
A similar plot for $n=1$ is given in \fig{f:NLLL}.
When solving \eq{LRdr} it turns out that there can be an imaginary part of
$E_h$ on-shell for certain values of $p_z$. Let us take the $R$
branch as an example. For $p_z\gg\cM$ the hole branch can exist if
$Es_{n-1}+p_zr_{n-1}$ does not grow with $p_z$, and that is possible since
$s(E,p_z^2+eB(2n-1))$ can be negative when the logarithm in
\eq{sandr} dominates, i.e. $E\simgeq\sqrt{p_z^2+eB(2n-1)}$.
The larger $p_z$ is, the closer $E$ is to the above square root.
This is why the hole branch goes exponentially close to the light cone
for large $p_z$ in
the standard analysis \cite{Weldon89}.
Here we also have a factor $(Es_{n}-p_zr_{n})$ in the right
dispersion relation in \eq{LRdr}. In $s_{n}$ and $r_{n}$ the argument
in the logarithms contains $E-\sqrt{p^2+eB(2n+1)}$. It is then clear that
the energy can go below the $n$-th light cone when $p_z$ is large enough
and the logarithm picks up an imaginary part.
The whole approximation
breaks down at these points since the excitations are no longer stable.
The value of the spectral function does not satisfy \eq{Zsum} when this
happens and it is not clear to what extent it is meaningful to continue
beyond these points, or even to go close to them.
%
% DP 95.06.26
%%%%%%%%%%%%%%%%%%%%%%%%%%%
%%%\Section{amm}{Energy in the lowest Landau level -- Anomalous magnetic moment}
\section{Energy in the lowest Landau level -- Anomalous magnetic moment}
\label{amm}
%%%%%%%%%%%%%%%%%%%%%%%%%%%%
The anomalous magnetic moment of electrons is defined from
the energy spectrum of a particle nearly  at rest in a weak field.
Usually, it is calculated from the triangle diagram
(see \fig{f:triangle}), using plane waves as
external states,
while it really should be done in the proper Landau levels as
explained in Section \ref{weakb}.
Furthermore, not only is the shift of one level required but
also the energy gap between different levels generated by the field.
In vacuum the anomalous energy shift is independent of the
Landau level for low enough fields (see e.g. \cite{GepragsRHRW94}).
We have not yet calculated the thermal shift for an arbitrary
\LL{},
but there are some interesting features already
in the lowest Landau level.
In vacuum one can calculate the \amm{} either from a triangle diagram,
using plane waves as external states, or extract it from the linear
term in the exact expression with the electrons in the general
Landau levels \cite{Schwinger51}.
The result is the same. At finite temperature
the issue is more complicated because of the IR sensitivity and
 it really makes a
difference if the external states are
in a proper \LL, or plane-wave states.
A related  problem was discovered in \cite{ElmforsS91,pebss91}, where a
real-time  method was used for the triangle
diagram, but we later found that the
limit procedure of taking the external photon
momentum to zero is not unique and thus the result is not
well defined. We shall exemplify this phenomenon here by
calculating the \amm{}  both from the expressions in Section \ref{weakb}
and from the triangle diagram (Section \ref{triangle}) in the static
limit.
\jump
%%%%%%%%%%%%%%%%%%%%%%%%%%%%%%%%%%%%%%%%%%%%%%%%%%
%%%\Subsection{selfamm}{The Self-energy method}
\section{The Self-energy method}
\label{selfamm}
%%%%%%%%%%%%%%%%%%%%%%%%%%%%%%%%%%%%%%%%%%%%%%%%%
Comparing with the energy shift
for a charge $-e$, particle with spin (${\bf s}$)
magnetic moment
$ \mbox{\boldmath $\mu$} =-e/2m \, g {\bf s} $ in a magnetic field
${\bf B}=-B {\bf e}_z$,
 $\dele=-\mbox{\boldmath $\mu$} \cdot {\bf B}$, we can write for an electron
 in the \LLL\ ( here $\la s_z \ra=  1/2$):
\be{ammdef}
        \dele \equiv \la \Psi_{1;0,0,0}|\hat{\Sigma}
        |\Psi_{1;0,0,0}\ra \equiv \dmbm -
        \frac{eB}{2m} \frac{\delta g}{2} +{\cal O}(eB)^2~~~,
\ee
where $\dmbm$ is the thermal mass, and $\delta g/2$ is the anomaly of the
magnetic moment. To linear order in the magnetic field, \eq{LLLvac} gives the
very well-known result \cite{Schwinger48,Schwinger51,kobsak83}
\be{delevac}
        \dele_{\rm vac}=- \frac{eB}{2m} \frac{\alpha}{2\pi}~~.
\ee
The total energy remains positive even for very large $B$ when
the non-linear terms dominate \cite{DittrichB81}.
 Let us now consider
 the thermal  photon contribution using Schwinger's pro\-per-time method.
 When the operators in
\eq{finph} act on  a \LL{}, we can use the properties
        (see Appendix \ref{app:efp})
\bea{eigvalLL}
        \Pi^2_\orto  \Psi^{(\pm)}_{\zeta;n,p_y,p_z}
        &=&eB[2n+\sigma_z]\Psi^{(\pm)}_{\zeta;n,p_y,p_z}\ ,\nn
        \sigma_3\Psi^{(\pm)}_{1;0,0,0}&=& \Psi^{(\pm)}_{1;0,0,0}\ ,\\
        \Pi\slask_\orto\Psi^{(\pm)}_{1;0,0,0}&=& 0\ ,\non
\eea
to compute the expectation value of $\hat{\Sigma}^\beta_\gamma$
in the \LLL. In the limit of vanishing $p_z$ we obtain to linear order in $eB$
\be{expvalseph}
       \dega(p_z=0) =
        m\left(\frac{\alpha\pi}{3}\frac{T^2}{m^2}
        +\frac{eB}{2m^2}\frac{2\alpha\pi}{9}\frac{T^2}{m^2}\right)\ ,
\ee
which agrees with standard results
\bort{\cite{FujimotoHY82,PeressuttiS82,DonoghueHR8485,JohanssonPS86}}
[29--32], and also with
the Furry picture in \eq{bzeroga}
and  the limit
$p_z \rightarrow 0$ in  \eq{blinga}. We also notice that
the main contribution comes from the hard part of the loop
integral ($k\simeq T$) and there is no
particular IR sensitivity.
%\bigskip

For thermal electrons we similarly obtain in the limit of vanishing $B$ and
$p_z$:
\bea{deemass}
        \align\dee(B=0,p_z=0) \equiv \dmbm _{e^+e^-}\nn
        &&= \frac{\alpha}{2\pi} \int_{-\infty}^\infty dk_0
        \Theta(k_0^2-m^2) f_F(k_0) \sqrt{k_0^2-m^2}
        \frac{k_0-2m}{m(k_0-m)}~~~,
\eea
which agrees with the corresponding limit of \eq{bzeroele}.
The thermal-electron contribution
to linear order in the magnetic field is more involved.
First, there is a $B$
dependence in the $\Theta$-function in \eq{finel}, which after
expansion generates a derivative of the distribution
function. Furthermore, one cannot expand the integrand in
$B$ naively, because that would generate IR-divergent integrals.
Another delicate integration by parts of the logarithm in \eq{finel}
is needed to give a
finite integral after the  expansion. There is thus  an ambiguity in the
way of writing
 the self-energy, depending on which terms are integrated by parts
to produce derivatives of the distribution function. We have chosen to
perform  further integrations by parts to remove the derivative on the
distribution function whenever this does not produce any IR divergences.
 After these manipulations
we find, up to linear order  in $B$:
\bea{expvalseel}
        \dee&=&\dmbm_{e^+e^-} +
        \frac{eB}{2m}\frac{\alpha}{3\pi} \int_m^\infty
        \frac{d\omega}{\sqrt{\omega^2-m^2}} \non \\
        && \times \left[ \left( \frac{2\omega^2+
        2m\omega-m^2}{m^2}-\frac{m}{(\omega+m)} \right)f_F^+(\omega)
         + 2m \frac{df_F^+(\omega)}{d\omega} \right]+ \non \\
        &&  +\frac{eB}{2m}\frac{\alpha}{3\pi} \int_m^\infty
        \frac{d\omega}{\sqrt{\omega^2-m^2}} \frac{ 2\om^3-3m^2\om -2m^3}
        {m^2(\om+m)} f_F^-(\om)~~~.
\eea
In order to obtain this result using the Furry picture propagator, we must
 take the limit $p_z \rightarrow 0$ in \eq{blinele}. Obviously the expansion
will be in powers of $p_z/(k_0-m)$. This  produces IR singularities, if we
are not careful. For the antiparticle part $-k_0 = \om>0$, so here it is
 straightforward to perform the expansion. For the particle part we
know, from above and by comparison with results using the imaginary-time
formalism, that it is likely that the finite result should contain
derivatives of the distribution function. Let us therefore introduce
the factor of convergence $(\om-m)^\nu$, where $\nu$ is assumed to be so
large that we may perform the expansion to get the
contribution for vanishing $p_z$. We may then  isolate
the term that will become
divergent as $\nu \rightarrow 0$. Integrating this term by parts,
the out-integrated term  vanishes for $\nu >1$. In the end we may consider
the analytical continuation $\nu \rightarrow 0$,
and arrive at \eq{expvalseel}.

In the high-temperature limit we obtain
\be{htseLLL}
       \dee \simeq
      m\left(  \frac{\alpha\pi}{6} \frac{T^2}{m^2}
        +\frac{eB}{2m^2}\frac{\alpha\pi}{9}\frac{T^2}{m^2}\right) .
\ee
Again, it agrees with standard results for the high-temperature limit
and it is dominated by a hard thermal loop. The difference with
\eq{expvalseph} is that \eq{htseLLL} is only approximate in the
high-temperature limit and that sub-leading terms {\it are} IR-sensitive.
This is why we had to be so careful with the expansion in $B$.
\input FEYNMAN
%%%%%%%%%%%%%%%%%%%%%%%%%%%%%%%%%%%%%%%%%%%%%%%
\Subsection{triangle}{The triangle diagram}
%%%%%%%%%%%%%%%%%%%%%%%%%%%%%%%%%%%%%%%%%%%%%%
The traditional way of computing the anomalous magnetic moment
is from the
triangle diagram in \fig{f:triangle}.
%
% ** Triangle **
\bfig
\begin{picture}(20000,12000)(-20000,-7000)
   \bigphotons
   \drawline\photon[\E\REG](-7000,0)[8]
      \drawarrow[\NE\ATBASE](\pmidx,\pmidy)
      \put(\pmidx,\pmidy){\makebox(300,-250)[tr]{$q$}}
      \put(\pbackx,\pbacky){\makebox(-150,900)[tr]{b}}
   \thicklines
   \drawline\fermion[\SE\REG](\photonbackx,\photonbacky)[4948]
      \drawarrow[\NW\ATBASE](\pmidx,\pmidy)
      \put(\pmidx,\pmidy){\makebox(-100,-100)[tr]{$k$}}
   \drawline\fermion[\NE\REG](\photonbackx,\photonbacky)[\fermionlength]
   \drawarrow[\NE\ATBASE](\pmidx,\pmidy)
      \put(\pmidx,\pmidy){\makebox(-700,800)[tr]{$k+q$}}
      \put(\pbackx,\pbacky){\makebox(-300,600)[tr]{a}}
   \thinlines
   \drawline\photon[\S\REG](\pbackx,\pbacky)[7]
      \global\advance\pmidx by -200
   \drawarrow[\S\ATBASE](\pmidx,\pmidy)
      \put(\pmidx,\pmidy){\makebox(3800,100)[tr]{$k-p$}}
      \put(\pbackx,\pbacky){\makebox(-300,-300)[tr]{c}}
   \thicklines
   \drawline\fermion[\NE\REG](\photonfrontx,\photonfronty)[\fermionlength]
   \drawarrow[\NE\ATBASE](\pmidx,\pmidy)
      \put(\pmidx,\pmidy){\makebox(1500,100)[tr]{$p'$}}
   \drawline\fermion[\SE\REG](\photonbackx,\photonbacky)[\fermionlength]
   \drawarrow[\NW\ATBASE](\pmidx,\pmidy)
      \put(\pmidx,\pmidy){\makebox(1500,200)[tr]{$p$}}
\end{picture}
\bc
\figcap{The triangle diagram.}
\label{f:triangle}
\ec
\efig
In vacuum, the full vertex, sandwiched between plane-wave states,
can be written as
\bea{Gcurrent}
        \bar{u}_{p'}\Gamma_\mu(p',p)u_p&=&
        \bar{u}_{p'}\left[F_1(q^2)\gamma_\mu+F_2(q^2)
        \frac{i\sigma_{\mu\nu}q^\nu}{2m}\right]u_p\nn[2mm]
        &=&
        \bar{u}_{p'}\left[\left(F_1(q^2)+F_2(q^2)\right)
        \gamma_\mu-F_2(q^2)
        \frac{p'_\mu+p_\mu}{2m}\right]u_p\ ,
\eea
where the Gordon decomposition was used in the last equality. The anomalous
magnetic moment is defined  by $\delta g/2=F_2(0)$ and
it can be extracted from the term linear in
$p_\mu$ in the limit $q\goto 0$. When using the Gordon decomposition,
it should be kept in mind that the external states are on the
$B=0$ mass shell. The use of such states is questionable from
the point of view of perturbation theory, as discussed in the beginning
of Section \ref{weakb}.

At finite
temperature there are two
formalisms for calculating the triangle diagram, the imaginary-
and real-time formalisms (ITF and RTF). In the ITF
we can put $q_\mu=0$ from the outset but we are left with
an analytic continuation in the $p_0$ variable.
On the other hand, in the RTF (we shall use
Thermo Field Dynamics (TFD) as a RTF) we get the result for
real $p_0$, but it is more tricky to put $q_\mu=0$ before
doing the loop integration, because there are potentially
ill-defined products of distributions.
There has recently been much interest in two issues that are
of importance here. First, the relation between ITF and TFD
has been clarified.
It was first found that the naive diagrams in TFD (the ones
with only physical fields on the external legs) did not
give the same result as the ITF in general
\cite{FujimotoMUO84,MatsumotoNU85,Evans87}. The difference was
shown to be related to the different kinds of analytic
continuations that are possible in the ITF, and different
time-ordered, retarded and advanced Green functions
\cite{Evans90,Kobes90,AurencheB92}. It is now clear that a certain
combination of TFD diagrams gives the same result as the
usual analytic continuation ($\omega\rightarrow p_0+i\epsilon$)
in ITF, which is the retarded Green function.
Secondly, the Lorentz invariance is broken at finite temperature
so the vertex function depends on $q_0$ and $\vek{q}$
independently. We can thus take the $q_\mu\rightarrow 0$
limit with either $q^2>0$ or $q^2<0$ (time-like and
space-like limits) or generally with any fixed ratio
$\abs{\vek{q}}/q_0$. It has been known for some time
that e.g. the one-loop self-energy in the $\phi^3$-theory
does not have a unique limit when $q_\mu\rightarrow 0$
\cite{FujimotoY88,GriboskyH90}.
In principle it would be interesting to discuss both limits,
but the external mass shell conditions on $p$ and $p'$
put the constraint on $q$ that
\be{constraint}
   p_0q_0-\vek{p}\vek{q}+\inv{2}(q_0^2-\vek{q}^2)=0\ ,
\ee
which prevents the use of $q_0$ and $\vek{q}$ as independent
variables. We shall therefore concentrate on the case
with $q\equiv 0$ for both the ITF and RTF.

In the ITF  we can simply put $q$ to zero from the outset.
We use standard ITF rules in Euclidean space and write the expression
for the vertex correction as
\be{vercorr}
        -ie\Lambda_\mu=(-ie)^3\int [d^4k] \frac{-i}{(k-p)^2+i\epsilon}
        \left(\frac{i}{k^2-m^2+i\epsilon}\right)^2
        T_\mu(k)\ .
\ee
The simplified tensor structure in the numerator is
\be{Tmu}
   T_\mu(k)\equiv\gamma^\nu(k\sslask+m)\gamma_\mu(k\sslask+m)\gamma_\nu=
   2(k^2-m^2)\gamma_\mu+4(2m-k\sslask)k_\mu
   \ .
\ee
The last term can be replaced by $4(2m-k_0)k_i$, since we are only interested
in terms proportional to $p_i$ (and not to $\gamma_i$) and
we neglect higher-order
terms in $p_i$ in the non-relativistic limit. Then also $\gamma_0$ can be
replaced by $\id$.  The double pole
in $k^2-m^2$
can conveniently be represented by a derivative,
\be{deriv}
   \at{\frac{\pa}{\pa M^2}\inv{k^2-M^2+i\epsilon}}{M^2=m^2}
   =\inv{(k^2-m^2+i\epsilon)^2}\ .
\ee
After an
analytic continuation we find as usual two contributions, one
from thermal photons and one from thermal electrons.
The photon contribution agrees with \eq{expvalseph} so we leave that
aside. From the thermal electrons we obtain
\bea{gfaktor}
        \frac{\delta g^{\beta,\mu}_{e^+e^-}}{2}&=&-\frac{\alpha}{3\pi}
        \int_m^\infty d\omega (\om^2-m^2)^{3/2}
        \left\{\frac{2}{m^2\om^2}(f^+_F(\om)+f^-_F(\om))
        \right.
        \nn[3mm]
        &&\left.-\left[\frac{df^+_F(\omega)}{d\om}
        \frac{2m-\om}{m\om(m-\om)^2}
        +\frac{df^-_F(\omega)}{d\om}
        \frac{2m+\om}{m\om(m+\om)^2}\right]\right\}\ .
\eea
We would  like to remark here  that special care has to be taken when
 splitting into matter and
vacuum contributions,
when the Feynman diagram considered contains fermion as well as boson
propagators.
In the so-called finite-density contribution (see e.g. Eq.~(3.71) in
 \cite{Kapusta89}),
 there is a term $\Theta(p_0-\om)$
that could have been mistaken for $\Theta(\mu-\om)$. This term has to be
added to the vacuum contribution in order to give the correct Feynman pole
prescription for the vacuum part of the fermion propagator.
The leading high-temperature contribution comes from the highest
power of $\om$ in the integral and it is given by
\be{htdg}
   \frac{\delta g^{\beta,\mu}_{e^+e^-}}{2}\simeq -\frac{\alpha \pi}{9}
   \left(\frac{T}{m}\right)^2\ ,
\ee
which agrees with the result in \eq{htseLLL}. The sub-leading terms,
which are sensitive to the IR part of the loop integral, are
on the other hand completely different.

In TFD there are several different vertex functions depending
on how the external legs are chosen to be particle or thermal
ghost lines and we call them $\Lambda_{abc}$ ($a,b,c=1,2$).
 According to \cite{Kobes90,Kobes91,AurencheB92} it is
necessary to sum over 1 and 2 with certain weights for all external
points except one. The external point with only a 1-field
is the one with the latest time argument. In our case we
consider an incoming electron going through the heat-bath
and interacting with an external magnetic field. The incoming
point $c$ in \fig{f:triangle}
must then have earlier time than point $a$. Also,
 the external field at point $b$ does not influence the
electron at later times than the time at point $a$. We are thus
naturally led to consider the combination with $a$ as the
latest time.
According to \cite{Kobes90} the retarded
Green function with the outgoing electron at the latest
time and $q=0$ is given by
\be{Lam}
        \Lambda=(\Lambda_{111}+\Lambda_{121})+
        e^{-\beta p_0/2}(\Lambda_{112}
        +\Lambda_{122})\ .
\ee
In \eq{Lam} there are several terms with overlapping $\delta$-functions
 for $q_\mu=0$ but they actually add up to give something finite. In particular
we see that the incoming photon line at point $b$
should be summed over 1 and 2 with
the consequence that there is no product of $\delta$-functions from
the two fermion propagators.  The $2\times 2$ structure of the
matrix propagator from $c$ to $a$ becomes
\be{ctoaprop}
   -T_\mu(k) \left(\ba{cc} G^2(k)-\sin^2\vartheta[G^2(k)-G^{*2}(k)]
                 & \sin\vartheta\cos\vartheta[(G^2(k)-G^{*2}(k)] \\[4mm]
                 -\sin\vartheta\cos\vartheta[(G^2(k)-G^{*2}(k)]
                 &G^{*2}(k)-\sin^2\vartheta[G^2(k)-G^{*2}(k)]\ea\right)\ ,
\ee
where $G(k)=(k^2-m^2+i\epsilon)^{-1}$ and $\sin^2\vartheta=
f_F(k_0)$.
The combination $G^2(k)-G^{*2}(k)$ turns
out to be related to a derivative of a $\delta$-function,
\be{derdel}
   G^2(k)-G^{*2}(k)=2\pi i\at{\frac{\pa}{\pa M^2}}{M^2=m^2}
   \delta(k^2-M^2)\ ,
\ee
which eventually leads to exactly the same expression as in the
ITF. The remaining multiple $\delta$-functions do not contribute for on-shell
external particles when $q_\mu=0$. The only thermal contributions are linear
in the electron and photon distribution functions. The same form
of regularization of the product between a propagator and a $\delta$-function
with the same argument was advocated in \cite{BedaqueD92}, but here it followed
simply from using the rules of \cite{Kobes90}.
It has been shown for the
$n$-point function in a scalar $\phi^3$-theory that the
ITF (and the mass-derivative formula) is recovered in
the space-like limit but {\it not} in the time-like \cite{FujimotoY88}.
This suggests that putting $q_\mu=0$ from start corresponds to the
space-like limit and the time-like limit could very well give
a different result. For a static external field, with space-dependent
gauge fields, it seems most appropriate to take the space-like limit.
Nevertheless, we do not find agreement between the triangle
diagram \eq{gfaktor} and $\delta g_{e^+e^-}^{\beta,\mu}$ inferred from
\eq{expvalseel}, apart from the leading high-temperature terms.
We surmise that the reason for this is that perturbation theory in $B$
is not adequate, as discussed at the beginning of this section,
and that the full Landau levels should be used as external states.
%
%%%%%%%%%%%%%%%%%%%%%%%%%%%%%%%%%%%%%%%%%%
%%%\Section{strongb}{The strong-field limit}
\section{The strong-field limit}
\label{strongb}
%%%%%%%%%%%%%%%%%%%%%%%%%%%%%%%%%%%%%%%%%%
In the limit of extremely strong magnetic fields,
the energies in all but the
 \LLL\
are approximately proportional to $\sqrt{eB}$.
 This implies that
intermediate states of higher \LL{}s are suppressed. It is thus very
convenient to use the Furry picture propagator with its explicit
spectral decomposition  in this case. One may then neglect the contributions
from all but  the \LLL , i.e. the sum over $n$ is reduced to $n=0$. In
Section \ref{selfstrong} we calculate the self-energy in this approximation.
In Section \ref{stronggauge} we explicitly verify that in this limit of
strong magnetic fields the self-energy is
independent of the gauge-fixing parameter in the photon propagator.
\jump
%%%%%%%%%%%%%%%%%%%%%%%%%%%%%%%%%%%%%%%%%%%%%%%%%%%%%%%%
\Subsection{selfstrong}{The self-energy in strong fields}
%%%%%%%%%%%%%%%%%%%%%%%%%%%%%%%%%%%%%%%%%%%%%%%%%%%%%%%%
In \cite{kobsak83} it was shown that the  approximation of only considering
intermediate electrons in the \LLL\  gives the same
result as the high-field limit of \eq{LLLvac}, i.e.
\be{hbvac}
        \dele_{\rm vac} \simeq \frac{m^2}{E_0}
         \frac{\alpha}{4\pi} \left(\ln
        \frac{2eB}{m^2} \right)^2 + {\cal O}\left(\ln\frac{2eB}{m^2}
         \right)~~,~~~ eB \gg m^2,p_z^2~~~.
\ee
When  the  contribution from thermal electrons is considered,
 it is perfectly clear from the
suppression by the distribution functions that the \LLL\ is dominating
for $eB \gg m^2, p_z^2, \mu^2, T^2$. After reducing
the sum to $n=0$ in \eq{LLLthe} and
performing the integrals as before, we may now also perform the
remaining Gaussian integrals
over $x,x'$. The result reads
\be{hbee}
        \dee  \simeq \frac{m^2}{E_0} \frac{2\alpha}{\pi^2} eB \int d^4k
        \frac{f_F(k_0) \delta(k_0^2-m^2-k_z^2)}{(k_0-E_0)^2-(k_z-p_z)^2-
        2eB k_\perp^2 +i \ve} \exp(-k_\perp^2)~~~.
\ee
Let us now use the $\delta$-function  to integrate over $k_z$. Then we
perform the
angular integral in ${\bf k}_\perp$ and substitute $u=k_\perp^2$.
We find
\bea{deeway}
        \lefteqn{\dee
        \simeq -\frac{m^2}{E_0}  \frac{\alpha}{2\pi}
        \int_{-\infty}^\infty dk_0
         \frac{\Theta(k_0^2-m^2)}{\sqrt{k_0^2-m^2}}f_F(k_0)
          }\non \\
        &&\times\sum_{\pm}
        \int_0^\infty du \frac{\exp(-u)}{u-[(k_0-E_0)^2-(p_z \pm
         \sqrt{k_0^2-m^2})^2+i\ve]/2eB}~~~.
\eea
We may here identify exponential
integrals, and use their asymptotic expansions
as given in Appendix \ref{app:expi}. The leading
strong-field behaviour is thus
\bea{deestrong}
        \dee&\simeq&-\frac{m^2}{E_0}
         \frac\alpha\pi \,  \int_{-\infty}^\infty dk_0
         \frac{\Theta(k_0^2-m^2)}{\sqrt{k_0^2-m^2}}f_F(k_0)
        \ln\left( \frac{eB}{m|k_0-E_0|} \right)  \non \\
        && -i \frac{m}{E_0}
        \alpha \int_m^\infty \frac{d\omega}{\sqrt{\omega^2-m^2}}
                f_F^-(\omega)~~~.
\eea
Notice here the field-independent (in this limit)
imaginary part, which  is exponentially suppressed for weaker
fields. We have also investigated the self-energy for a positron in the \LLL .
The result is the same as in the electron case above, but $f_F^-(\omega)$
is replaced by $f_F^+(\omega)$ in the imaginary part on the right-hand side of
\eq{deestrong}.
The physical process which is responsible for the
occurrence of the imaginary part of $\dee$ is the reaction
$e^+\,e^-\mapsto\gamma$, which is possible in the presence of an external
field that may absorb momentum to make the process energetically possible
 \cite{Wunner79}. We have calculated in Appendix~\ref{app-epa} the
tree-level decay rate $\Gamma$, for an electron in  a
background of positrons. The result is just as expected,
$\Gamma \equiv 2\,\Im \dee$.
The way we obtain the imaginary part in $\dee$ in the strong-field limit
with the $+\ve$ prescription in \eq{deeway} is similar to the pole
prescription of Landau in the theory of longitudinal plasma oscillations
(see e.g. \cite{Oraevsky83}).

For the thermal-photon contribution we use instead the $\delta$-function
to integrate over $k_\perp$, and find the result
corresponding to \eq{deeway},
\be{degaway}
        \dega  \simeq \frac{m^2}{E_0}  \frac{\alpha}{\pi}
        \int_{-\infty}^\infty dk_0
        f_B(k_0) \int_{-k+p_z}^{k+p_z}dk_z\,  \frac{
        \exp\{-[k^2-(k_z-p_z)^2]/2eB\} }
        {k_z^2- (k^2+2E_0 k_0+p_z^2)-i\ve}~~~,
\ee
where $k\equiv |k_0|$. We may, to leading order, approximate the exponential
function with unity.
The poles of the integrand are outside the interval of integration in $k_z$,
so the thermal photon contribution is real, and we may let $\ve$ vanish.
However, $ k_0^2+2E_0 k_0+p_z^2$  changes sign at $k_0=-E_0 \pm m$, so we
must split the $k_0$ integral in two parts.
After performing the $k_z$-integral we thus obtain
\bea{degastrong}
        \lefteqn{\dega \simeq \frac{m^2}{E_0}
         \frac\alpha\pi \int_{E_0-m}^{E_0+m}
                \frac{dk f_B(k)}{\sqrt{2 E_0 k -k^2-p_z^2}} } \non \\
        &&\times \left[ \arctan\left( \frac{k+p_z}{
        \sqrt{2 E_0 k-k^2-p_z^2}} \right) + \arctan\left( \frac{k-p_z}{
        \sqrt{2 E_0 k-k^2-p_z^2}} \right) \right] \non \\
        &&-\frac{m^2}{E_0} \frac\alpha{2\pi} \left\{ \int_{-\infty}^{-E_0-m}+
        \int_{-E_0+m}^\infty \right\}
        \frac{dk_0 f_B(k_0)}{\sqrt{k^2+2E_0 k_0+p_z^2}}
        \nn
        &&\times\ln \left( \frac{k^2+ E_0 k_0+k \sqrt{k^2+2E_0 k_0+p_z^2}}
        {k^2+E_0 k_0 -k \sqrt{k^2+2E_0 k_0+p_z^2}} \right). \non \\
        &&
\eea
Notice that this is finite since there are cancellations
between all of the  ostensible IR  divergences. For $p_z^2 >0$, $k_0=0$
is contained in the last integral in \eq{degastrong}. To leading order the
integrand is odd in $k_0$ for small $k_0$, and a cancellation  thus occurs.
The next leading term  produces a finite result when integrated over $k_0$.
For $p_z=0$ there is a cancellation between the leading terms from the first
and last integrals in \eq{degastrong}, and the result is still finite.

We have plotted the different contributions to the self-energy in the
high-field limit $\{eB\gg m^2,~p_z^2,~\mu^2,~T^2 \}$: in
\fig{fig-bfhigh} as a function of the magnetic field, in
\fig{fig-tehigh} as a function of the temperature, and in
\fig{fig-pzhigh} as a function of the momentum in the $z$-direction
 parallel to the magnetic field. The self-energy is even in $p_z$.
In each case the self-energy is small compared to the electron rest mass and
 thus will only provide a small energy shift. In the high-field limit it is
the vacuum contribution that dominates. Using \eq{hbvac} we see that for
the self-energy to be of the same order of magnitude as the electron mass
we must have $eB/m^2\simeq 10^{17}$. This corresponds to the mind-bogglingly
large field of $B\simeq 10^{27} $ T.
\begin{figure}[tbp]
\centerline{ \psfig{figure=dispbfpl,height=8cm}}
\vspace{6ex}
\figcap{The electron self-energy $\dele=\dele_{\rm vac}+\dee+\dega$ in the
high-field limit $ \{eB\gg m^2,~p_z^2,~\mu^2,~T^2 \}$,
as a function of the magnetic field,
for $p_z/m=1,~\mu/m=1$, and $T/m=1$. }
  \nopagebreak
  \label{fig-bfhigh}
\end{figure}
\begin{figure}[tbp]
\centerline{ \psfig{figure=disptepl,height=8cm}}
\vspace{6ex}
\figcap{The electron self-energy $\dele=\dele_{\rm vac}+\dee+\dega$ in the
high-field limit $\{eB\gg m^2,~p_z^2,~\mu^2,~T^2 \}$,
as a function of the temperature,
for $eB/m^2=8,~p_z/m=1$, and $\mu/m=1$. }
  \nopagebreak
  \label{fig-tehigh}
\end{figure}
\begin{figure}[tbp]
\centerline{ \psfig{figure=disppzpl,height=8cm}}
\vspace{6ex}
\figcap{The electron self-energy $\dele=\dele_{\rm vac}+\dee+\dega$ in the
high-field limit $\{eB\gg m^2,~p_z^2,~\mu^2,~T^2 \}$
as a function of the momentum parallel to the magnetic field,
for $eB/m^2=8,~\mu/m=1$, and $T/m=1$. Each contribution is even in $p_z$.}
  \nopagebreak
  \label{fig-pzhigh}
\end{figure}
%
%%%%%%%%%%%%%%%%%%%%%%%%%%%%%%%%%%%%%%%%%%%%%%%%%%%%%
\Subsection{stronggauge}{The gauge-fixing dependence}
%%%%%%%%%%%%%%%%%%%%%%%%%%%%%%%%%%%%%%%%%%%%%%%%%%%%
 We formally show in Appendix
\ref{app:gauge} that the self-energy is independent of the gauge-fixing
parameter $\xi$ on the tree-level mass shell. In the limit of extremely
strong fields we may explicitly verify this conclusion.
We shall here consider the different contributions to the self-energy appearing
for non-vanishing $\xi$ in \eq{photprop}. To regularize the vacuum contribution
we assume the external electron to be off-shell
\be{offenerg}
        E^2-m^2-p_z^2\equiv \Delta~~~,
\ee
and we shall find that the gauge-dependent part is proportional to
$\Delta$, as shown formally in \cite{KobesKR91}.
Factorizing out  $\xi$, the vacuum contribution reads
\bea{xivac}
        \lefteqn{\xi \dele_{{\rm vac},\xi}(B\rightarrow \infty) \simeq  \xi \,
        i \frac{e^2}{(2\pi)^3} \frac{2eB}{2m(E+p_z)} \int_{-\infty}^\infty
         dk_0 \, dk_z
        \int_0^\infty du e^{-u} }\non \\
        && \times \Bigl\{ (k_0+k_z)\Delta^2+ (k_0^2-m^2-k_z^2+i\ve)
        [-2E^3-(E^2-m^2)k_z+ \non \\
        &&+(E^2+m^2)k_0 + 2 p_z\{E(k_0-k_z)-E^2+p_z^2\} +
        p_z^2(2E+k_0-k_z)]\Bigr\}
        \non \\
        &&\times \frac{1}
        {(k_0^2-m^2-k_z^2+i\ve)[(k_0-E)^2-(k_z-p_z)^2-2eBu+i\ve]}~~~.
\eea
Let us now use the Feynman parametrization
\be{trifeyn}
        \frac{1}{a^2b}=2\int_0^1 ds \frac{ s}{[(1-s)b+sa]^3}~~~,
\ee
in the first term on the right-hand side of \eq{xivac}. We may then
integrate over $k_0$ and $k_z$, to obtain
\bea{xivacii}
        \lefteqn{\dele_{{\rm vac},\xi}(B\rightarrow \infty) \simeq
        \frac{\Delta}{m} \frac\alpha{4\pi} \int_0^\infty du e^{-u}\,
        \left\{ \frac1{u-i\ve} + \right.} \non \\
        &&+ \left. 2eB \Delta \,
         \int_0^1 ds \frac{s^2}{[m^2(1-s)^2-\Delta s(1-s) +2eB u s-i\ve]^2}
                \right\}~~~.
\eea
Notice that we cannot naively let $\Delta=0$ here, since the result would  be
an ill-defined product of $\Delta$ and a logarithmic divergence.
The $u$ integral is dominated by small $u$. The $s$ integral would be dominated
by $s\simeq 1$. Inserting $s=1$, except in the terms $(s-1)$, we may
easily perform the $s$ integral. The result will contain simple
 polynomials and logarithms. Expanding the logarithms for $u m^2/eB\ll 1$ to
leading order,
several cancellations will occur. The result reads
\be{xivacagain}
        \dele_{{\rm vac},\xi}(B\rightarrow \infty) \simeq -\frac\alpha{4\pi}
        \frac{\Delta^2}{m(2m^2-\Delta)} \int_0^\infty du \frac{e^{-u}}
        {u +(m^2-\Delta)/2eB-i\ve}~~~.
\ee
We may here identify an exponential integral, and use its asymptotic expansion
to get
\be{xivacres}
        \dele_{{\rm vac},\xi}(B\rightarrow \infty) \simeq -\frac\alpha{4\pi}
        \frac{\Delta^2}{m(2m^2-\Delta)} \, \ln \! \left( \frac{2eB}{m^2-\Delta}
        \right)~~~.
\ee
This is obviously vanishing on the tree-level mass shell $\Delta=0$.
However, notice that if we would like to solve the dispersion relation
self-consistently in this limit, the gauge-fixing
 dependence is only logarithmically
suppressed compared to the result in the Feynman gauge, proportional to
$[\ln(2eB/m^2)]^2$ according to \eq{hbvac}.
When considering the contribution of thermal electrons to the $\xi$ dependence,
the integrand in correspondence to \eq{xivac} is proportional to
\be{gaugele}
        \delta(k_0^2-m^2-k_z^2)\{(k_0+k_z)\Delta^2 +(k_0^2-m^2-k_z^2)
        [-2E_0^3 +  \ldots]\}~~~.
\ee
This is vanishing on the tree-level mass shell $\Delta=0$.
Also for thermal photons we may immediately use the on-shell energy.
The contribution to the self-energy multiplied by $\xi$ then becomes
\be{xigamma}
        \dele_{\gamma,\xi}^\beta
        \simeq \frac\alpha\pi eB \frac{E_0}m \int_0^\infty du\, e^{-u}\,
        \int_{-\infty}^\infty dk_0 \, dk_z \, k_0
        \delta'(k_0^2-k_z^2-2eB u) f_B(k_0)~~~,
\ee
where the prime on the $\delta$ function
denotes a derivative with respect to its argument.
Since the photon is its own antiparticle, $f_B(k_0)$ has to be even in $k_0$.
Then $\dele_{\gamma,\xi}^\beta$ is vanishing due to the antisymmetric
 integration in $k_0$.
%% PE 95.05.05
%%%%%%%%%%%%%%%%%%%%%%%%%%%%%%%%%%%%%%%%%%%%%
%%%\Section{disc}{Discussion and final remarks}
\section{Discussion and final remarks}
\label{disc}
%%%%%%%%%%%%%%%%%%%%%%%%%%%%%%%%%%%%%%%%%%%%%
%
In this paper we have presented a detailed account on the
fermion self-energy when both an external magnetic field and
the presence of a heat bath have to be taken into account. In the
context of astrophysics, many QED processes in the presence of
strong magnetic fields have been studied, however mostly without  the
presence of a heat bath.

As was mentioned in Appendix \ref{app:ternov}
it has been argued in the literature
that one must use a Dirac spinor basis which   diagonalizes in the
Sokolov--Ternov spin operator \cite{SokolovT68,TernovD94}
 in order to consistently treat unstable
excited relativistic Landau levels in perturbation theory
\cite{Graziani93,GrazianiHS95}. We have
verified that such a basis can be constructed from the Dirac spinor basis
defined in  Appendix \ref{app:efp}.
We have also seen that when a heat bath is
present the situation is more complex and that in general such a basis does
not diagonalize the self-energy
 except in the case of zero momentum  parallel to the
external magnetic field.

The effect of thermal quasi-particles on the emission of neutrinos
of a very hot or dense star, but without the presence of a magnetic
field, has   been studied in
\cite{Braaten92}. It was found that under certain
physical conditions the electron--plasmino annihilation may exceed
the electron--positron process into neutrinos. In principle it is
possible to carry out a similar analysis in the presence of a
magnetic field. As we have seen in Section \ref{ht}, the presence of
a magnetic field leads to a $p_z$ asymmetric quasi-particle spectrum.
We have suggested that   this asymmetry may play a role also in
dynamics, leading to the observed high space velocities of pulsars
\cite{Chugai84,DorofeevRT85}.

We have also carried out a detailed analysis of the anomaly of the
magnetic moment $\delta g/2$ when a thermal heat bath of photons,
electrons and positrons is present. Previous considerations
\cite{ElmforsS91,pebss91}
of this problem have led to ill-defined IR behaviour. With thermal
photons, electrons and positrons
present, we obtain the following contribution to the
anomaly:
\bea{geeg}
         \frac{\delta g^{\beta , \mu}_{e^+e^{-},\gamma}}{2} &=&
         - \frac{2\alpha \pi}{9} \frac{T^2}{m^2} \nn
        && -\frac{\alpha}{3\pi} \int_m^\infty
        \frac{d\omega}{\sqrt{\omega^2-m^2}} \non \\
        && \times \left[ \left( \frac{2\omega^2+
        2m\omega-m^2}{m^2}-\frac{m}{(\omega+m)} \right)f_F^+(\omega)
         + 2m \frac{df_F^+(\omega)}{d\omega} \right]+ \non \\
        &&  + \frac{\alpha}{3\pi} \int_m^\infty
        \frac{d\omega}{\sqrt{\omega^2-m^2}} \frac{ 2\om^3-3m^2\om -2m^3}
        {m^2(\om+m)} f_F^-(\om)~~~.
\eea
To the  best of our knowledge this is the first IR well-defined expression for
the anomalous magnetic moment when a thermal environment of photons
and electrons is present. At zero temperature the $\om$-integral in \eq{geeg}
can be carried out explicitly, and we find
\be{dgTnoll}
        \frac{\delta g^{\beta ,\mu}_{e^+e^{-}}}{2}=\
        -\frac{\alpha}{3\pi}\left[
        \left(2+\frac{\mu}{m}-\frac{m}{m+\mu}\right)
        \sqrt{\left(\frac{\mu}{m}\right)^2-1}-
        \frac{2m}{\sqrt{\mu^2-m^2}}\right]~~,
\ee
for $\mu>m$. In the  limit of large $\mu$, this agrees with the result in
\cite{ElmforsS91}, as could be expected since the IR sensitivity
is sub-dominant.
Thus the conclusion in \cite{ElmforsS91} that $g$ can become considerably
smaller than 2 for high densities remains valid. Using a bold extrapolation to
large corrections we find that $g+\delta g\simeq 0$ for $\mu\simeq 35\,m_e$,
which should be compared with a typical chemical potential $\mu\simeq 300\,m_e$
inside a neutron star. Even if the approximations are not valid for such a
large $\mu$ there could still be important corrections to the synchrotron
radiation from the surface of a neutron star, and thus to the estimation of the
$B$-field from observations.

We notice also in \eq{dgTnoll} that $\delta g$ seems to become very large for
small densities, e.g. when $\mu\goto m$. This is an artefact of the expansion
in $B$. As discussed in Section \ref{weakb} we do not expect an expansion in
$B$ to be universally possible, but some sort of de~Haas--van~Alphen
oscillations should  show up in certain limits. For $T=0$ and $B$ fixed we can
take the limit $\mu\goto m$ directly in \eq{LLLthe} and it corresponds actually
to a strong-field calculation since only the \LLL{} is inside the Fermi sea.
The leading contribution can then be extracted from \eq{deeway}
using \eq{expiserie} and we obtain
\be{delowdens}
        \Delta E=-\frac{\alpha}{\pi}\sqrt{\mu^2-m^2}~
        \left[\ln\left(\frac{2eB}{\mu^2-m^2}\right)
          +2-C\right]+\cO\left(\frac{\mu^2-m^2}{eB}\right)~~,
\ee
valid for $m<\mu<\sqrt{m^2+eB}$.
Our conclusion is that the energy shift is perfectly finite when $\mu\goto m$
(in fact, it goes to zero), but since $\Delta E$ does not admit a power series
expansion in $B$ the standard definition of a magnetic moment
(see \eq{ammdef}) does not make sense.

In a strong magnetic field we have, finally, observed the presence of an
 imaginary part in the self-energy which  becomes
field-independent if the external magnetic field is large enough. Formally,
this imaginary part appears in a way similar to the pole description of
Landau in the theory of longitudinal plasma oscillations \cite{Oraevsky83}.
This imaginary
contribution is therefore   a relativistic counterpart of Landau damping and
is physically due to the possibility of $e^{+}e^{-}$ annihilation  into a
{\it single } photon in a magnetic field \cite{Wunner79}.
\vspace{13mm}
\begin{center}
{\bf ACKNOWLEDGEMENT}
\end{center}
\vspace{3mm}
One of the authors (B-S. S.) wishes to thank
Gabriele Veneziano and the TH Division for
the hospitality at CERN when this work was
finalized. P.~E. and B-S.~S. thank Benny Lautrup
and Kimmo Kainulainen for
discussions. B-S.~S. also acknowledges support in part by the
Swedish National Research Council under contract No. 8244-316 and
the Research Council of Norway under contract No. 420.95/004.
\vspace{3mm}
%
%
%% PE 95.06.16
\jump
\appendix
\begin{center}
        {\Large{\sc APPENDICES }}
\end{center}
\Section{app:efp}{External-field propagator}
In this appendix we summarize, for the convenience of the reader,
the relevant
expressions used in the main text for a constant magnetic field $B$ in
the negative  $z-$direction in the gauge $A_\mu=(0,0,Bx,0)$. We use
the $\gamma$-matrices   in the chiral representation,
\be{chiralgam}
        \gamma_{0} = \left( \begin{array}{cc}
        0 & -\id \\
        -\id & 0
        \end{array} \right)~~~,~~~~~\gamma ^{i} = \left(  \begin{array}{cc}
        0 & \sigma^i \\
        -\sigma^i & 0
        \end{array} \right)~~~.
\ee
We seek solutions to the
Dirac equation for a  fermion  in an external field
\be{dirac}
        (i\not\!\!D - m)\Psi^{(\pm)}_\kappa ({\bf x},t) = 0~~~,
\ee
where $iD\slask=\gamma^\mu(i\pa_\mu+eA_\mu)$ for a particle of charge $-e$.
With our choice of gauge it follows that
\be{psiform}
  \Psi^{(\pm)}_{\kappa}({\bf x},t)= \frac{1}{2\pi\sqrt{2E}}
        \, \exp[ \pm i(
  - E t \plus  p_{y}y \plus  p_{z}z ) ] \,
    \Phi^{(\pm)}_{\kappa}(x)~~~,
\ee
where $\kappa$ denotes $p_y,~p_z$, and any other quantum-number necessary
to specify the wave functions. For ease of notation  we shall write
 $iD_\mu \rightarrow \Pi_\mu=
        (E,-p_z,-\mbox{\boldmath $\Pi$}_\perp) $ when
acting on the above wave functions.
Acting with $(\Pi\slask + m)$ on \eq{dirac}, using
$[\Pi_x,\Pi_y]=ieF_{xy}=ieB$,
 we find
\be{squaredir}
        ( E^2-m^2-p_z^2- \Pi_\perp^2+eB \sigma_{xy})\Psi=0~~~,
\ee
where $\sigma_{xy}\equiv i [\gamma_x,\gamma_y]/2 =
{\rm diag}[\sigma_z,\sigma_z]$.
Let us now introduce the functions
\bea{Indef}
  I_{n;p_{y}}(x)& \equiv& \left( \frac{eB}{\pi} \right)^{1/4} \exp \left[
  - \frac{1}{2} eB \left( x \minus \frac{p_{y}}{eB} \right)^{2} \right]
 \frac{1}{ \sqrt{n!}} H_{n} \left[ \sqrt{2eB} \left( x \minus \frac{p_{y}}
 {eB} \right) \right] ~~~ ,\nonumber \\
 &&
\eea
where  $H_{n}$ is the Hermite polynomial given by the Rodrigues formula as
\be{hermite}
  H_{n}(x)=(-1)^{n} e^{\frac{1}{2}x^{2} } \frac{d^{n}}{dx^{n}}
e^{- \frac{1}{2}
            x^{2}} ~~~ ,
\ee
and we define $ I_{-1;p_{y}}(x)=0$.
The functions $ I_{n;p_{y}}(x)$ are normalized according to
\be{Innorm}
\int dx I_{n;p_{y}}(x) I_{n';p_{y}}(x) = \delta _{n,n'}~~~,
\ee
for $ n,n'=0,1,2, \ldots$~. Defining $\xi_\pm=\Pi_x\mp i\Pi_y$,
we have
$\Pi_\orto^2=\xi_+\xi_- +eB$, and they
act on the $I_n$ functions according to
\be{xirel}
        \ba{l}
        \xi_+ I_{n-1;p_y}(x)=i\sqrt{2eBn} I_{n;p_y}(x)\ , \\[3mm]
        \xi_- I_{n;p_y}(x)=-i\sqrt{2eBn} I_{n-1;p_y}(x)\ .
        \ea
\ee
It readily follows that
\be{pioni}
        (\Pi_\perp^2-eB \sigma_z)\, \mbox{\rm diag}
        [I_{n;p_y}(x),I_{n-1;p_y}(x)]=
        2eBn\, \mbox{\rm diag}[I_{n;p_y}(x),I_{n-1;p_y}(x)]~~~.
\ee
We may therefore write $\Phi$ in \eq{psiform} in the form
\be{phiform}
        \Phi^{(\pm)}_{\zeta;n,p_y,p_z }(x)=
        \mbox{\rm diag}[I_{n;p_y}(x),I_{n-1;p_y}(x),I_{n;p_y}(x),
        I_{n-1;p_y}(x)]u^{(\pm)}_{\zeta,n,p_y,p_z}~~~,
\ee
where $u^{(\pm)}_\kappa$ is a Dirac spinor independent of $x_\mu$, and
 $\zeta=\pm 1$ denotes a polarization index.
The energy eigenvalues are given by the relativistic \LL s
\be{LLenergy}
 E= E_{n}(p_z) = \sqrt{m^{2} + p_z^{2} +2eBn}~~~.
\ee
In the \LLL\ $(n=0)$, $I_{-1;p_y}\equiv 0$ implies that there is only one
choice of $u_\kappa$ possible, corresponding to $\zeta=1$.
 For the higher \LL{}s there is a twofold degeneracy, since two
linearly
independent $u^{(\pm)}_{\zeta,n,p_y,p_z}$ may be found,
and there is thus an ambiguity
in the choice of wave functions.

 In \cite{Parle87} there is a discussion about a choice of
the $u^{(\pm)}_{\zeta,n,p_y,p_z}$ which diagonalizes the self-energy
 corrections in
vacuum. We have not found any simple generalization of that basis at finite
temperature, as stated in Appendix \ref{app:ternov}.
 On the other hand, in this paper we  need no such basis
since we always deal either with the expectation value of the
self-energy in the
non-degenerate lowest Landau level or with the full
$4\times 4$ matrix. For any higher Landau levels it is, of course,
mandatory to
diagonalize the matrix including corrections since the perturbation theory is
degenerate, as emphasized e.g. in \cite{GepragsRHRW94}.

We shall here use the choice of  wave functions given in \cite{kobsak83}
\be{phidef1}
  \Phi^{(+)}_{1;n,p_y,p_z}(x) =\frac{1}{\sqrt{E_n+p_z} }
    \left( \ba{c}
             (E_{n} \plus p_{z} ) I_{n;p_{y}}(x) \\
             - i \sqrt{2eBn}\, I_{n-1;p_{y}}(x) \\
             - m I_{n;p_{y}}(x) \\
              0
          \ea \right)~~~ .
\ee
\be{phi2}
 \Phi^{(+)}_{-1;n,p_y,p_z}(x) =\frac{1}{\sqrt{E_n+p_z } }
    \left( \ba{c}
             0 \\
             - m I_{n-1;p_{y}}(x) \\
             - i \sqrt{2eBn}\, I_{n;p_{y}}(x) \\
             ( E_{n} \plus p_{z} ) I_{n-1;p_{y}} (x)
           \ea \right)~~~ ,
\ee
\be{phi3}
   \Phi^{(-)}_{1;n,p_y,p_z}(x) =\frac{1}{\sqrt{E_n-p_z} }
    \left( \ba{c}
           -m I_{n;-p_{y}}(x) \\
           0 \\
           (-E_{n} \plus p_{z} ) I_{n;-p_{y}}(x) \\
           i \sqrt{2eBn} \,  I_{n-1;-p_{y}}(x)
          \ea  \right)~~~ ,
\ee
\be{phi4}
  \Phi^{(-)}_{-1;n,p_y,p_z}(x) =\frac{1}{\sqrt{E_n-p_z } }
    \left( \ba{c}
           i \sqrt{2eBn} \,  I_{n;-p_{y}}(x) \\
            (-E_{n} \plus p_{z} ) I_{n-1;-p_{y}}(x) \\
             0 \\
             - m  I_{n-1;-p_{y}}(x)
           \ea \right) ~~~ .
\ee
 It can be  shown that the collection of
all $\Psi$'s forms a complete orthonormal set. The wave functions are
normalized according to
\be{psinorm}
        \int d^3 x \Psi^{(\lambda)\dagger}_{\zeta;n,p_y,p_z}(x)
        \Psi^{(\lambda')}_{\zeta';n',p'_y,p'_z}(x)=\delta_{\zeta,\zeta'}\,
        \delta_{n,n'}\,\delta_{\lambda,\lambda'}\,
        \delta(p_y-p'_y)\,\delta(p_z-p'_z)~~~,
\ee
where $\lambda=\pm$.
The  propagator in the external field $ iS(x',x)\equiv iS_{\rm vac(x',x)}+i
S^{\beta,\mu}(x',x)$, as given in Eqs.~(\ref{propform}),\,(\ref{termprop}),
is then explicitly written (see e.g. \cite{kobsak83} for the vacuum part,
 and \cite{ElmforsPS94} for the additional term when we have fermions
according to some one-particle distribution):
\bea{bvprop}
      S(x',x)_{ab}&= &\sum^{\infty}_{n=0}\int
       \frac{dk_0 \, dk_{y} \, dk_{z}}{(2\pi)^{3}} \exp[-ik_0 (t' -t) +
        ik_{y}(y' - y) + ik_{z}(z' - z)] \nonumber \\
         & & \times \left[ \frac{1}{k_0^{2}
         \minus k_{z}^{2} \minus m^{2} \minus 2eBn  +i\ve}
        + 2 \pi i \delta(k_0^{2}
       \minus k_{z}^{2} \minus m^{2} \minus 2eBn)f_F(k_0)\right]\nonumber \\
         &&\times S_{ab}(n;k_0, k_{y}, k_{z};x',x) ~~~,
\eea
where
\be{fFdef}
 f_F(k_0) = \Theta(k_0 )f_{F}^{+}(k_0) \plus \Theta(-k_0 ) f^{-}_{F}(-k_0)~~~,
\ee
 The matrix $S(n;k_0 , k_{y}, k_{z},x',x)$ entering above  is in the chiral
representation of $\gamma$-matrices explicitly written
\bea{bprop}
  \lefteqn{~~~~~~~~~~~~~~~~~~~~~~~~~~~~~~~S(n;k_0 , k_{y}, k_{z}) \equiv}
 \nonumber
\\[2ex]
 && \left( \ba{cccc}
     mI_{n,n} & 0 & - (k_0 \plus k_{z}) I_{n,n} & - i \sqrt{2eBn}
                  I_{n,n-1} \\
    0 & m I_{n-1,n-1} & i \sqrt{2eBn} I_{n-1,n} & - (k_0 \minus k_{z})
                  I_{n-1,n-1} \\
    - (k_0 \minus k_{z} ) I_{n,n} & i \sqrt{2eBn} I_{n,n-1} &
       m I_{n,n} & 0 \\
    - i \sqrt{2eBn} I_{n-1,n} & - (k_0 \plus k_{z} ) I_{n-1,n-1} &
       0 & m I_{n-1,n-1}
    \ea \right)~~~ ,\nonumber \\
\eea
where we have used the short-hand notation
\be{InIn}
  I_{n',n} \equiv I_{n';k_{y}}(x') I_{n;k_{y}}(x)~~~~.
\ee
%================================================================
%%%\Section{app:ternov}{The Sokolov--Ternov spin operator}
\section{The Sokolov--Ternov spin operator}
\label{app:ternov}
Let $H_D=\mbox{\boldmath $\alpha$}\cdot\pv+\beta m$, where
$\alpha^i=\gamma^0\gamma^i,~\beta=\gamma^0$,
be the Landau--Dirac Hamiltonian in an external magnetic field. Let the
 magnetic field be parallel to the $z$-axis. The $z$-component of
the Sokolov--Ternov \cite{SokolovT68} spin operator is
$\mhz=\sigma_z(m+\Pi\slask_\orto)$. (For a recent discussion of this
operator in the context of calculations of the electron anomalous magnetic
moment, see \cite{TernovD94}.) It follows that $\mhz$ is conserved,
i.e. it commutes with $H_D$. It is easy to construct linear combinations
of the solutions in \eq{psiform}, which also diagonalize $\mhz$.
Suppressing all the quantum numbers of the solutions in \eq{psiform}
except the polarization $\zeta=\pm1$, solutions to the Dirac equation
(\ref{dirac}) that diagonalize $\mhz$ can be written in the form
\be{Psimhz}
   \Psi^{(\pm)}_\mhz=N_\pm(\Psi_+-ia_\pm\Psi_-)\ ,
\ee
where, if $n\neq 0$,
\be{aN}
      a_\pm=\frac{p_n}{m\pm E_n(0)}\ ;
      \quad N_\pm=\frac{p_n^2}{p_n^2+(m\pm E_n(0))^2}\ ,
\ee
where $p_n=\sqrt{2eBn}$. The states in \eq{Psimhz} have the property
$\mhz\Psi^{(\pm)}_\mhz=\pm\sqrt{m^2+p_n^2}\Psi^{(\pm)}_\mhz$.
The \LLL{} is diagonal in $\mhz$ with eigenvalue $m$. It can be verified
that the propagator $S(x',x)$, when
calculated in the basis in \eq{Psimhz}, still is the same as the one
given in Appendix \ref{app:efp}. The basis in \eq{Psimhz} furthermore
has the property that the self-energy  $\Sigma(x',x)$ is diagonal in
the vacuum sector to $\cO(\alpha)$.
By using the fact that the on-shell Sokolov--Ternov
operator can be written in the form $\mhz=\sigma_z(\gamma^0p^0-\gamma_zp_z)$,
this follows easily from the fact that $[\mhz,\beta\hat{\Sigma}_0]=0$,
where $\hat{\Sigma}_0$ is the vacuum self-energy operator \cite{Tsai74}.
In the presence of a thermal heat bath the self-energy operator
$\beta\hat{\Sigma}^{\beta,\mu}$, defined as the sum of \eq{thphseop}
and \eq{thelseop}, commutes with $\mhz$ only if $p_z=0$.
It has been observed \cite{Graziani93,GrazianiHS95}
that in the vacuum sector one must use the basis where $\mhz$ is diagonal
in order to introduce  consistently in perturbation theory, a resonance
width into the QED cyclotron scattering amplitudes.
In a thermal environment the situation is more complex
and further studies are required.
%==================================================================
%%%\Section{app:tadpole}{The tadpole}
\section{The tadpole}
\label{app:tadpole}
Also the tadpole could
 possibly contribute to the one-loop self-energy.
 In configuration space the tadpole is proportional to
\be{tad}
        J^\nu(x)\equiv \tr[e\gamma^\nu iS(x,x)]~~~,
\ee
where the trace is over spinor indices.
Introducing an $\ve >0$ in  the argument of $\Psibar~:t'
\rightarrow t+\ve$, we may remove the time-ordering. Then we let
$\ve$ vanish and immediately find
\be{current}
        J^\nu(x)=-e \la \Psibar(x)\gamma^\nu \Psi(x)\ra~~~,
\ee
i.e. the expectation value of the electromagnetic current.
We could instead choose to  change the argument of $\Psi:~t
\rightarrow t+\ve$,
in order to remove the time ordering, and then let $\ve$ vanish.
In this case we need
to use
the equal time anticommutation relation $\{\Psi_a({\bf x'},t),
\Psi^\dagger_b({\bf x},t)\}_+=\delta_{ab}\delta^3({\bf x'}- {\bf x})$
to reverse the order of $\Psibar$ and $\Psi$. Since $\gamma^\nu$ is traceless,
we again arrive at \eq{current}.  Let
us now  separate the
current in its vacuum and thermal contributions $J^\nu=J^\nu_{\rm vac}+
J^\nu_{\beta,\mu}$. Trivially $J^\nu_{\rm vac}$ vanishes after
renormalization,
and so does $J^j_{\beta,\mu}~$, for $j=1,2,3$ in our static model.
What is left is the charge density $J^0_{\beta,\mu}$. But in order for our
static model to be valid there must be no such charge asymmetry on average.
If there is a finite chemical potential for electrons
there must be  a compensating (static) background
charge to make the average charge density vanish everywhere.
Actually, it is easily shown with explicit calculations for
the particular system we consider that the
only possible
 non-vanishing part of $J^\nu$, as defined in \eq{tad}, is $J^0_{\beta,\mu}$.
Using the electron
propagator in \eq{bvprop} and its thermal generalization, it follows that
$J^x$ and $J^z$ vanish when performing the trace. Also $J^y$ and
$J^0_{\rm vac}$
vanish due to the antisymmetric integration in $k_y$ and $k_0$, respectively.
In the  case of vanishing chemical potential, $f_F(k_0)$ is even, and then
also $J^0_{\beta,\mu}$ vanishes due to the antisymmetric integration.
In the case of general $\mu$ we find
\be{chargedens}
        J^0_{\beta,\mu}= -e \,\frac{eB}{(2\pi)^2} \sum_{n=0}^\infty \int dk_z
        \{f_F^{+}[E_n(k_z)]-f_F^{-}[E_n(k_z)]\}(2-\delta_{n,0})~~~,
\ee
i.e. the charge density. Notice the factor
$(2-\delta_{n,0})$, which originates in a term $I_{n,n}+I_{n-1,n-1}$, and
shows the twofold degeneracy in all but the lowest \LL{}s.
%
%
%================================================================
%%%\Section{app:gauge}{Gauge independence}
\section{Gauge independence}
\label{app:gauge}
We have used the Feynman gauge throughout the article and now we shall show
that the result is gauge-independent on-shell, i.e. ~ $\langle \Psi_\k|
\hat{\Sigma} |\Psi_{\k'}\rangle$ is independent of the gauge-fixing parameter
$\xi$ when $\Psi_\k$ and $\Psi_{\k'}$ are solutions to the Dirac equation.
The only
possible $\xi$ dependence in a general covariant gauge comes from the photon
propagator. In vacuum, it would give a term of the form
\be{gdep}
        \xi\int d^4x'\, d^4x \Psibar_\k(x') ie^2\gamma^\mu
         [\pa_\mu \pa_\nu' \tilde{D}(x'-x)] S(x,x')\gamma^\nu
        \Psi_{\k'}(x)\ ,
\ee
where
\be{Dtilde}
        \tilde{D}(x'-x)=\int [d^4k]\frac{e^{-ik(x'-x)}}{(k^2+i\epsilon)^2}\ .
\ee
After one partial integration in $x$ or $x'$ the Dirac equation can be used
together with $(iD\slask-m)S(x,x')=\delta(x-x')$ to show the \eq{gdep} is zero.
At finite temperature the gauge-dependent part of the photon propagator is
again given by a total derivative \cite{Landsman86} and the same proof goes
through.
%
%================================================================
%%%\Section{app:expi}{Exponential integrals}
\section{Exponential integrals}
\label{app:expi}
In \cite{hmf} we have the following definitions of the  exponential integrals
\bea{expint}
        {\rm E}_1(z) &\equiv & \int_z^\infty dt \frac{e^{-t}}t~~~,~|\arg z|<
        \pi~~~, \\
        {\rm Ei}(x)&\equiv & - {\cal P}\int_{-x}^\infty dt\,
        \frac{e^{-t}}t~~~,~x \in \real~~~.
\eea
Furthermore they satisfy the interrelation
\be{inter}
        {\rm E}_1(-x\pm i\ve) =-{\rm Ei}(x) \mp i \pi~~~,~x>0~~~,
\ee
and have the series expansions
\bea{expiserie}
        {\rm E}_1(z)&=&-C-\ln z -\sum_{n=1}^\infty
        \frac{(-1)^n z^n}{n n!}~~~,~|\arg z|< \pi~~~, \\
        {\rm Ei}(x)&=&C+\ln x +\sum_{n=1}^\infty
        \frac{x^n}{n n!}~~~,~x \in \real~~~,
\eea
where $C=0.5772156649\ldots$ is Euler's constant.
%
%================================================================
%%%\Section{app-epa}{Electron-positron annihilation}
\section{Electron-positron annihilation}
\label{app-epa}
%+++++++++++++++++++++++++++++++++++++++
We shall here consider the process $e^- e^+ \mapsto \gamma$,
where the positron comes from the  heat and charge bath.
A large magnetic field may absorb momentum to make the process energetically
possible. Consider an electron in the state described by $\Psi^{(+)}_p$,
a positron  in the state described by $\Psi^{(-)}_\kappa$, and
a photon with momentum $q_\mu$ and  polarization $\ve_\mu(\lambda,q)$.
 To the lowest order in perturbation theory (i.e. perturbatively
in the quantum electromagnetic field, the static uniform magnetic field
is treated exactly), we find the corresponding transition matrix element
\be{trans}
        i T_{p,\kappa;q,\lambda}=-ie \int d^4 x \Psibar_\kappa^{(-)}(x)\,
        \gamma^\mu \Psi^{(+)}_p(x) \ve^*_\mu(q,\lambda) e^{i q \cdot x}~~~.
\ee
Let us now form $| T_{p,\kappa;q,\lambda}|^2$. Sum over all polarizations
of the outgoing photon
\be{photpol}
        \sum_\lambda \frac{\ve^*_\mu(q,\lambda)\,\ve_\nu(q,\lambda)}
        {\ve^*(q,\lambda)\cdot \ve(q,\lambda)}=g^{\mu\nu}~~~,
\ee
and integrate over the photon  momentum
\be{photint}
        \left. \int\frac{ [d^3 {\bf q}]}{2 q_0} \right|_{q_0=|{\bf q}|}=
        \int [d^4 q]\, 2\pi\, \delta(q_0^2-{\bf q}^2) \, \Theta(q_0)~~~.
\ee
Sum over all incoming positrons with distribution $f_{F}^{-}(E_\kappa)$.
Due to the normalization of our wave functions in \eq{psinorm}
we must divide by $\int dy\,dz$, for the norm of the decaying electron state.
  Dividing by the infinite time elapsed
between the final and initial state $\int dt$, we find the decay rate
\bea{decay}
        \Gamma&=&\frac{e^2}{\int dy\, dz\, dt } \int d^4x\int d^4 x' \,
        \Psibar^{(+)}_{\zeta;n,p_y,p_z }(x) \, \gamma^\mu \sum_{\kappa}
        f_F^-(E_\kappa)
        \Psi_\kappa^{(-)}(x') \Psibar_\kappa^{(-)}(x) \gamma_\mu\non \\
        && \times \int [d^4 q] \,2\pi\, \delta(q^2) \Theta(q_0)
        e^{-i q \cdot(x'-x)}\,
        \Psi^{(+)}_{\zeta;n,p_y,p_z }(x')~~~.
\eea
By comparing with \eq{termprop} we see that $ \sum_{\kappa}
        f_F^-(E_\kappa)\Psi_\kappa^{(-)}(x') \Psibar_\kappa^{(-)}(x)$ is
exactly the thermal positron contribution to the Dirac fermion propagator.
Due to energy conservation (as follows when performing the integration
over $t$) $q_0<0$ will  not contribute, so we may drop $ \Theta(q_0)$.
The decay rate is thus  exactly equal to the imaginary part of the electron
self-energy,  with thermal positrons in the  intermediate states, obtained
through $1/(q^2+i\ve)={\cal P}(1/q^2) -i \pi \delta(q^2)$ in the photon
propagator. This splitting into principal and imaginary part is equivalent to
what was done in the high-field limit in Section~\ref{selfstrong}, using the
definition and interrelation of exponential integrals as given in
Appendix~\ref{app:expi}.
%================================================================
%
% PE 95.06.09
\jump

\end{document}